%% file: paper.tex
\documentclass[a4paper,fleqn,usenatbib]{mnras}

\DeclareMathAlphabet{\altmathcal}{OMS}{cmsy}{m}{n}
\usepackage{mathptmx}

\usepackage[T1]{fontenc}
\usepackage{ae,aecompl}

\usepackage{graphicx}	
\usepackage{amsmath}	
\usepackage{amssymb}	
\usepackage{xcolor}     
\usepackage{cancel}
\usepackage{multirow}
\usepackage{mathrsfs}
\usepackage{ulem}

\usepackage[T1]{fontenc}
\usepackage{ae,aecompl}

\newcommand{\fpath}{./Figs/}
\renewcommand{\vec}[1]{\mathbf{#1}}
\newcommand{\hvec}[1]{\hat{\boldsymbol{#1}}}
\newcommand{\pd}[2]{\frac{\partial #1}{\partial #2} }
\newcommand{\HALF}{\frac{1}{2}}

\newcommand{\quotes}[1]{``#1''}
\newcommand{\DS}{\displaystyle}
\newcommand{\tens}[1]{\mathsf{#1}}

\newcommand{\cF}{\altmathcal{F}}

\newcommand{\cU}{\altmathcal{U}}
\newcommand{\cV}{\altmathcal{V}}

\DeclareMathOperator{\sign}{sign}

\title{A comparison of approximate non-linear Riemann solvers for Relativistic MHD}

\author[G. Mattia]{G. Mattia$^{1}$\thanks{E-mail: {mattia@mpia.de}},
 A. Mignone$^{2,3}$
 \\
$^1$ Max Planck Institute for Astronomy and IMPRS -- University of Heidelberg, 
K\"onigstuhl 17, D-69117, Heidelberg Germany  \\
$^2$ Dipartimento di Fisica, Universit\`a di Torino, via P. Giuria 1, I-10125
Torino, Italy \\
$^3$ INAF, Osservatorio Astronomico di Torino, Strada Osservatorio 20, I-10025 Pino
Torinese, Italy \\
}

\date{Accepted XXX. Received YYY; in original form ZZZ}

\pubyear{2021}

\begin{document}
\label{firstpage}
\pagerange{\pageref{firstpage}--\pageref{lastpage}}
\maketitle

\begin{abstract}
We compare a particular selection of approximate solutions of the Riemann problem in the context of ideal relativistic magnetohydrodynamics. 
In particular, we focus on Riemann solvers not requiring a full eigenvector structure.
Such solvers recover the solution of the Riemann problem by solving a simplified or reduced set of jump conditions, whose level of complexity depends on the intermediate modes that are included.
Five different approaches - namely the HLL, HLLC, HLLD, HLLEM and GFORCE schemes - are compared in terms of accuracy and robustness against one- and multi-dimensional standard numerical benchmarks.
Our results demonstrate that - for weak or moderate magnetizations - the HLLD Riemann solver yields the most accurate results, followed by HLLC solver(s).
The GFORCE approach provides a valid alternative to the HLL solver being less dissipative and equally robust for strongly magnetized environments.
Finally, our tests show that the HLLEM Riemann solver is not cost-effective in improving the accuracy of the solution and reducing the numerical dissipation.
\end{abstract}

\begin{keywords}
methods: numerical -- relativistic processes -- MHD -- shock waves -- Riemann solver
\end{keywords}



\input{introduction.tex} 
\input{method.tex}
\input{results.tex}

\input{summary.tex}

\section*{Data Availability}

The PLUTO code is publicly available and the simulation data will be shared on reasonable request to the corresponding author.

\section*{Acknowledgments}

We acknowledge the helpful comments by an anonymous referee which helped to improve the quality
of this manuscript.

\bibliographystyle{mnras}
\bibliography{paper}

\appendix
\input{appendix.tex}

\bsp	
\label{lastpage}
\end{document}

%% file: introduction.tex
\section{Introduction}
%
%
%

Several astrophysical phenomena, such as jets and accretion flows around compact objects, gamma-ray bursts and pulsar wind nebulae, are closely connected to relativistic flows.
Because of the high non-linearity of the Relativistic MagnetoHydroDynamics (RMHD) equations, a numerical approach is unavoidable in order to expand our theoretical understanding of such relativistic phenomena.
In this regard, Godunov schemes have become the standard approach in the solution of hyperbolic conservation laws because of their built-in numerical viscosity and their ability to accurately capture discontinuous waves (such as shock waves).
In such methods, the discretization process heavily relies on the integral form of the equations so that conservation of mass, momentum and energy is naturally ensured.
A fundamental step of these shock-capturing schemes is the solution of the so-called Riemann problem, i.e. the decay of two separated and spatially constant states, determines the fluxes of the conserved quantities at each interface.
Unfortunately, due to its huge computational cost, an exact Riemann solver \citep{GR2006} is not a feasible option to solve a multidimensional problem.
Instead, approximate methods of solution are commonly preferred.

Over the last decades, several approximate solutions to the Riemann problem have been developed in the context of relativistic MHD.
Roe's type Riemann solvers (\citealt{Komissarov1999,Balsara2001,Koldoba_etal2002}) are based on the exact linearization of the equations and require the full characteristic decomposition.
Unfortunately, as pointed in \citet{ERMS1991} and \citet{Komissarov1999}, linear solvers may not satisfy the entropy condition through strong rarefactions.
In RMHD, a state of art of the Roe-type Riemann solvers has been developed by \citet{Anton_etal2010} \citep[and earlier by][]{Koldoba_etal2002}, which have  provided the (quite lengthy) analytical expressions for both right and left eigenvectors.
Albeit the linearized approach of Roe is capable of accounting for all the seven waves present in the solution, we shall not consider it here because of its heavy numerical cost.
For this reason we prefer to focus on incomplete Riemann solvers, which do not include in their structure the full set of waves.

A second family of (approximate) Riemann solvers (of which the HLL solver can be considered the progenitor) dates back to the original work of  \cite{HLL1983}.
The HLL Riemann solver has become extremely popular because of its ease of implementation, reduced computational cost and robustness \cite[see, e.g.][in the context of Special and General relativistic MHD]{Gammie2003,dZ_etal2007,BeckStone2011,WhiSto2016}).
The HLL scheme approximates only two out of the seven waves by collapsing the full structure of the Riemann fan into a single average state. 
Because of this, the solver has large numerical dissipation and has pushed the quest for more accurate approaches.

An extension of the HLL scheme, able to restore the contact wave, was developed originally by \citet{TSS1994} for the Euler equation.
The so-called HLLC (where 'C' stands for Contact) formulation, was later extended to RMHD by \citep{MB2006,HonJan2007}.
In both \citet{MB2006} and \citet{HonJan2007} the solution method differs depending on whether the normal component of the magnetic field vanishes or not.
A solution to this problem was brought by \citet{KimBal2014} and then improved in \citet{BalKim2016} who developed a HLLC solver which retrieves naturally the hydrodynamical limit when the magnetic field tends to zero.

A further step was made by \citet{MUB2009} who developed a HLL-type Riemann solver able to preserve both contact discontinuities and Alfv\'en waves by extending the classical solver of \citet{MiyKus2005} to relativistic MHD.
Despite it complexity, the HLLD (here 'D' stands for Discontinuities) is able to reduce drastically the numerical dissipation at the cost of solving a nonlinear equation through an iterative scheme.

Other approaches have also been attempted as well to restore the intermediate missing waves in the solution to the Riemann problem.
Following the approach of \citet{ERMS1991}, \citet{DumbserBalsara2016} have proposed a solution to the Riemann problem based on the HLLEM (called also HLLI in some papers) formulation which restores selected anti-diffusive flux terms on top of the HLL structure, in order to capture selected intermediate waves.

Finally, we also consider here the  generalized First Order Centered (GFORCE) scheme, originally formulated by \citet{TorTit2006} and recently employed by \citet{MigZan2021} in the context of Upwind Constrained Transport schemes for MHD.
The GFORCE flux comes as a weighted average of the Lax-Friedriechs and Lax-Wendroff fluxes and has reduced numeric dissipation when compared to the former.
It only requires the maximum characteristic wave speed.

The main goal of this paper is to provide an extensive quantitative comparison of the aforementioned Riemann Solvers in the context of ideal relativistic MHD.
Numerical tests in 1,2 and 3 dimensions are performed in order to assess computational speed, robustness and accuracy of the Riemann solvers mentioned above.
The main capabilities of each approach are documented, providing clear recipes about which is the most suited Riemann Solver depending on the context.

Our paper is structured as follows. 
In Section \ref{sec::Equations} we briefly describe the RMHD equations. 
In Section \ref{sec::Riemann} we describe the Riemann solvers studied in the paper.
In Section \ref{sec::Tests} we test the Riemann Solver through several numerical benchmarks.
Conclusions are finally drawn in Section \ref{sec::Summary}.

%% file: method.tex
\section{Equations of Ideal Relativistic MHD}
\label{sec::Equations}
%
%
%

We consider an ideal relativistic magnetized fluid \citep{Lichnerowicz1976,Anile2005} in flat space-time (with Minkowski metric tensor $\eta^{\mu\nu} = \text{diag}(-1,1,1,1)$)) described by the conservation of mass,
\begin{equation}
\label{eq::rest_mass_cov}
 \partial_\mu(\rho u^\mu) = 0,
\end{equation}
energy-momentum,
\begin{equation}
 \partial_\mu[(\rho h + b^2)u^\mu u^\nu - b^\mu b^\nu + p\eta^{\mu\nu}] = 0,
\end{equation}
and the Maxwell dual tensor,
\begin{equation}
\label{eq::maxwell_cov}
 \partial_\mu(u^\mu b^\nu - u^\nu b^\mu) = 0.
\end{equation}
Here we follow the standard convention that Latin indices take values for spatial components while Greek indices label space and time components.
The quantities introduced in equations (\ref{eq::rest_mass_cov})-(\ref{eq::maxwell_cov}) are, respectively, the fluid rest mas density $\rho$, the four-velocity $u^\mu$, the relativistic specific enthalpy $h$, the covariant magnetic field $b^\mu$ and the total pressure (thermal + magnetic) $p = p_g+|b^2|/2$.
Note that, in our units, the speed of light $c=1$ and a factor $\sqrt{4\pi}$ has been reabsorbed in the definition of $b^\mu$.
The four-vector $u^\mu$ and the fluid velocity $v^i$ are related through
\begin{equation}
u^\mu = \gamma\;(1,v^i),
\end{equation}
where $\gamma = (1 - v^2)^{-1/2}$ is the Lorentz factor, while the relation between $b^\mu$ and the laboratory magnetic field $B^i$ is
\begin{equation}
 b^\mu = \gamma[{\vec v}\cdot{\vec B},\frac{B^i}{\gamma^2} + v^i({\vec v}\cdot{\vec B})].
\end{equation}
The square modulus of the covariant magnetic field can be written as
\begin{equation}
 b^2 = \frac{B^2}{\gamma^2} + ({\vec v}\cdot{\vec B})^2 \,.
\end{equation}
The system of RMHD equation is closed through an appropriate equation of state.
Throughout the paper we assume an ideal gas equation of state, described by a constant
$\Gamma-$law
\begin{equation}
  h = 1 + \frac{\Gamma}{\Gamma - 1}\frac{p_g}{\rho},
\end{equation}
where $\Gamma$ is the adiabatic exponent, although alternative equations,
as in \citet{MMcK2007}, may be adopted.

The system of equations (\ref{eq::rest_mass_cov})-(\ref{eq::maxwell_cov}) can be written in the standard conservation form
\begin{equation}\label{eq::set_eq}
  \pd{\cU}{t} + \sum_{k}\pd{{\cF}^k}{x^k} = 0,
\end{equation}
(where $k = x,y,z$) together with the divergence-free condition of magnetic field
\begin{equation}
  \nabla\cdot{\vec B} = 0.
\end{equation}
The conserved variables and the fluxes along the direction $k$ are, respectively,
\begin{equation}\label{eq::flux}
 \cU = \left( \begin{array}{c}
  D \\
  m^i \\
  B^i \\  
  E
\end{array}\right)
 \,,
\qquad \cF^k = \left(
\begin{array}{c}
  Dv^k \\
  m^iv^k + p\delta^{ik} - b^iB^k/\gamma \\
  v^kB^i - v^iB^k \\
  m^k - Dv^k
\end{array}\right),
\end{equation}
where the quantities $D$, $m^i$ and $E$ stand, respectively, for the laboratory mass  density, the momentum density and the energy density (net of mass contribution).

In addition to the conserved variables $\cU$, the set of primitive variables $\cV = (\rho,v^i,B^i,p_g)$ is also routinely employed.
While the conversion from primitive to conserved variables can be recovered analytically through
\begin{equation}
\label{eq::C2P}
\begin{array}{lcl}
  D   & = & \rho\gamma \,  
  \\ \noalign{\medskip}
  m^i & = & (\rho h\gamma^2 + B^2)v^i - ({\vec v}\cdot{\vec B})B^i\, 
  \\ \noalign{\medskip}
  E   & = & \rho h\gamma^2 - p_g - \rho\gamma + \DS\frac{B^2}{2} +  \DS\frac{v^2B^2 - ({\vec v}\cdot{\vec B})^2}{2} \,,
\end{array}
\end{equation}
primitive variables must be computed numerically from the conserved quantities (see,  e.g., \citealt{dZBL2003,NGMCDZ2006,MMcK2007}).

From now on we assume a one-dimensional problem along the $x$-direction.
As for the non-relativistic case, the one-dimensional system of the RMHD equations involves 7 equations (since in 1D, the normal component of $\vec{B}$ is constant).
Integrating Eq. (\ref{eq::set_eq}) over the $i-$th cell and over a time step $\Delta t$, we get
\begin{equation}\label{eq:1Dblock}
 \cU^{n+1}_i = \cU^n_i - \frac{\Delta t}{\Delta x}
                \left(\hat{\cF}_{i+\HALF} - \hat{\cF}_{i-\HALF}\right),
\end{equation}
where $\Delta x$ is the mesh spacing and $\hat{\cF}$ is the numerical flux function which follows from the solution of a Riemann problem at zone interfaces where fluid quantities experience the discontinuity
\begin{equation} \label{eq:RiemannIC}
 \cU(x,0) = \left\{
 \begin{array}{cc}
 \cU_{L,i+\HALF} & \quad{\rm if} \quad x < x_{i+\HALF} \\ \noalign{\medskip}
 \cU_{R,i+\HALF} & \quad{\rm if} \quad x > x_{i+\HALF} \,. 
 \end{array}\right.
\end{equation}
Here $\cU_{L,i+\HALF}$ and $\cU_{R,i+\HALF}$ are, respectively, the left and right state values on either side of the zone interface $i+\HALF$.

The decay of the initial discontinuity defined by Eq. (\ref{eq:RiemannIC}) spawns a self-similar pattern comprised of seven waves, as in classical MHD \citep{Komissarov1999}.
At the double end of the Riemann fan, two fast magnetosonic waves bound the emerging pattern enclosing a pair of rotational (or Alfv{\'e}n) discontinuities, a pair of slow magnetosonic waves and a contact (or tangential) discontinuity in the middle. 
Fast and slow magnetosonic disturbances can be either shocks or rarefaction waves, depending on the pressure jump and the norm of the magnetic field. 
Primitive variables experience a jump across a fast or a slow shock, whereas thermodynamic quantities like thermal pressure and rest density remain continuous when crossing a relativistic Alfv{\'e}n wave. 
Contrary to its classical counterpart, however, the tangential components of magnetic field trace ellipses instead of circles and the normal component of the velocity is no longer continuous across a rotational discontinuity \citep{Komissarov1997}. 
Finally, through the contact mode, only density exhibits a jump while thermal pressure, velocity and magnetic field remain continuous.

\section{Non-linear, Approximated Riemann Solvers}
\label{sec::Riemann}
%

\begin{figure*}
\centering   
 \includegraphics[width=0.95\textwidth]{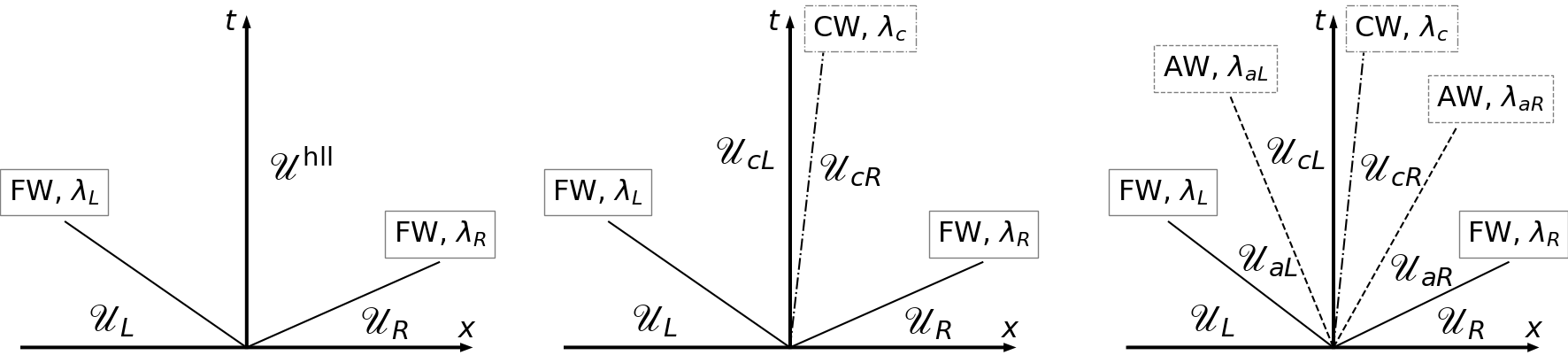}
 \footnotesize
 \caption{Riemann fan structure for the HLL, HLLC and HLLD approaches, respectively.}
 \normalsize
 \label{fig:riemann_fan}
\end{figure*}

\subsection{HLL Formulation}
\label{sec:HLL}
%

The HLL Riemann solver, originally devised by \cite{HLL1983} for the equations of gas-dynamics  \citep{dZBL2002,dZBL2003}, approximates the internal structure of the Riemann fan with a single state $U^{\rm hll}$ bounded by two outermost fast magnetotosonic waves (leftmost panel in Fig. \ref{fig:riemann_fan}).
This single state is required to satisfy the jump conditions across each of the two waves
\begin{equation}\label{eq:hll_jump}
  \begin{array}{l} 
  \lambda_L (\cU_L - \cU^{\rm hll}) =  \cF_L - \cF^{\rm hll}  \,, \\ \noalign{\medskip}
  \lambda_R (\cU_R - \cU^{\rm hll}) =  \cF_R - \cF^{\rm hll}   \,.
  \end{array}
\end{equation}
As such, the HLL approach avoid the full characteristic decomposition of the equations since only an estimate to the two outermost fast waves $\lambda_L$ and $\lambda_R$ is needed. 

Eqns. (\ref{eq:hll_jump}) yield a total of $14$ equations in the $14$ unknowns given by the components of $\cU^{\rm hll}$ and $\cF^{\rm hll}$ (note that $\cF^{\rm hll} \ne \cF^x(\cU^{\rm hll})$).
The solution is readily found as
\begin{equation}\label{eq:Uhll}
 \cU^{\rm hll} = \DS\frac{\lambda_R\cU_R - \lambda_L\cU_L + \cF_L - \cF_R}{\lambda_R - \lambda_L},
\end{equation}
and
\begin{equation}\label{eq:Fhll}
 \cF^{\rm hll} = \frac{\lambda_R\cF_L - \lambda_L\cF_R +
  \lambda_R\lambda_L(\cU_R - \cU_L)}{\lambda_R - \lambda_L},
\end{equation}
where $\cF_s = \cF^x(\cU_s)$, for $s = L,R$.
Eq. (\ref{eq:Uhll}) is also known as the integral representation of the Riemann fan \citep{Toro_1997}

The outermost wave speeds $\lambda_L$ and $\lambda_R$ represent an upper bound to the actual wave speeds and can be estimated using the initial left and right input states \cite[see, e.g][]{MB2006}.

The actual numerical flux is finally computed as follows:
\begin{equation}\label{eq:HLL_flux}
 \hat{\cF} = \left\{ \begin{array}{ll}
  \cF_L         & \quad {\rm if}\quad  \lambda_L \geq 0 \,, 
   \\ \noalign{\medskip} 
  \cF^{\rm hll} & \quad {\rm if}\quad  \lambda_L \leq 0 \leq \lambda_R\,,  
   \\  \noalign{\medskip}
  \cF_R         & \quad {\rm if}\quad  \lambda_R \leq 0\,.
\end{array}
\right.
\end{equation}

The HLL approach is simple to implement, cost-effective and requires only a guess to the outermost fast speed without any particular knowledge of the solution.
The major drawback, however, is its inability to resolve contact or tangential waves

\subsection{HLLC Formulation}
%

The HLLC formulation was originally proposed by \citet{TSS1994} and later extended to relativistic hydrodynamics by \cite{MB2005} and to relativistic MHD by \cite{MB2006, HonJan2007, KimBal2014}. 
The solver attempts to restore the intermediate contact wave thus leading to a two-state representation of the internal Riemann fan structure:
\begin{equation}
 \cU(x,t) = \left\{
 \begin{array}{ll}
  \cU_L   & \quad {\rm if}\quad \lambda_L \geq x/t, 
    \\  \noalign{\medskip}
  \cU_L^* & \quad {\rm if}\quad \lambda_L \leq x/t \leq \lambda^*, 
    \\  \noalign{\medskip} 
  \cU_R^* & \quad {\rm if}\quad \lambda^* \leq x/t \leq \lambda_R, 
    \\ \noalign{\medskip}  
  \cU_R  & \quad {\rm if}\quad \lambda_R \leq x/t \,,
 \end{array}\right.
\end{equation}
where $\lambda^*$ is now the velocity of the middle contact wave, see also the middle panel in Fig. \ref{fig:riemann_fan}.
Likewise, the corresponding numerical fluxes at the interface evaluates as:
\begin{equation}
 \hat{\cF}(0,t)  = \left\{
\begin{array}{ll}
  \cF_L   & \quad {\rm if}\quad  \lambda_L \geq 0, 
    \\ \noalign{\medskip}
  \cF_L^* & \quad {\rm if}\quad  \lambda_L \leq 0 \leq \lambda^*, 
      \\ \noalign{\medskip}
  \cF_R^* & \quad {\rm if}\quad  \lambda^* \leq 0 \leq \lambda_R, 
      \\ \noalign{\medskip}
  \cF_R   & \quad {\rm if}\quad \lambda_R \leq 0 \,.
\end{array}
\right.
\end{equation}
Intermediate states and fluxes must satisfy the Rankine-Hugoniot jump conditions:
\begin{equation}\label{eq::RH_HLLC}
 \begin{array}{cll}
    \lambda_L(\cU^*_L - \cU_L)   & = & \cF^*_L - \cF_L   \,,
      \\ \noalign{\smallskip}
    \lambda^*(\cU^*_R - \cU^*_L) & = & \cF^*_R - \cF^*_L  \,,
      \\ \noalign{\smallskip}
    \lambda_R(\cU_R   - \cU^*_R) & = & \cF_R   - \cF^*_R \,.
 \end{array}
\end{equation}

Adding together the previous equations yields the consistency condition
\begin{equation}\label{eq::consstateHLLC}
    \DS\frac{(\lambda^* - \lambda_L)\cU^*_L + 
    (\lambda_R - \lambda^*)\cU^*_R}
             {\lambda_R - \lambda_L} = \cU^{\rm hll} \,,
\end{equation}
or, upon dividing by the corresponding $\lambda$, the equivalent condition on the fluxes:
\begin{equation}
\label{eq::consfluxHLLC}
    \DS\frac{\cF^*_L\lambda_R(\lambda^* - \lambda_L) + 
             \cF^*_R\lambda_L(\lambda_R - \lambda^*)}{\lambda_R - \lambda_R}
     = \lambda^* \cF^{\rm hll} \,.
\end{equation}
In general one can not take $\cF^* = \cF(\cU^*)$ since fewer waves in the Riemann fan are accounted for.
For this reason we can look at Eqns. (\ref{eq::RH_HLLC}) as providing, in principle, $3\times7=21$ relations across three waves and a consistent solution can therefore be sought by introducing $21$ unknowns.
However, if the speed of the contact mode is chosen to coincide with the fluid normal velocity, the continuity equation across the middle wave is trivially satisfied and the number of equations reduces to $20$ ($10$ per state).
This allows states ($\cU^{*}_{L/R}$) and fluxes ($\cF^{*}_{L/R}$) in the star region to be expressed in terms of the $20$ unknowns
\begin{equation}
  \left(D, v_x, v_y, v_z, m_y, m_z, B_y, B_z, p, E\right)^*_{L/R} \,,
\end{equation}
with the condition $\lambda^* = v^*_{x,L} = v^*_{x,R}$.
The normal component of the momentum is not considered an independent quantity since it can be expressed through a combination of the previous unknowns as
\begin{equation}
  m^*_{x} = (E' + p)^*v^*_x - (\vec{v}\cdot\vec{B})^*B_x  \,,
\end{equation}
where $E' = E+D$, which holds both for the left or the right state in the \emph{star} region.
Note also that $B_x$ enters as a constant parameter in the solution process.

The HLLC solvers of \cite{MB2006,HonJan2007, KimBal2014} are based on this formalism although they require different conditions to be satisfied across the middle contact wave.
In the following we describe the original approach of \cite{MB2006} and the more recent improvement by \cite{KimBal2014}. 

\subsubsection{Solution of \citet{MB2006}}
\label{sec:HLLC_MB}
%

In the approach of \citet{MB2006} (henceforth HLLC-MB), the solution of the Riemann problem differs depending on whether the normal magnetic field vanishes or not.
When $B_x\ne 0$, the following conditions across the contact discontinuity are assumed:
\begin{equation}\label{eq:MB06_assumption}
    \begin{array}{lll}
    v^*_{x,L} = v^*_{x,R} \qquad & v^*_{y,L} = v^*_{y,R} \qquad & v^*_{z,L} = v^*_{z,R} \\ \noalign{\medskip}
    p^*_L = p^*_R \qquad & B^*_{y,L} = B^*_{y,R} \qquad & B^*_{z,L} = B^*_{z,R} 
    \end{array}
\end{equation}
The solution of the Riemann problem can then be divided into the following steps:
\begin{itemize}
    
\item[(i)] By virtue of their continuity, the transverse components of $\vec{B}$ are given by the HLL single state
    
\begin{equation}\label{eq:Bhll}
 B^*_y = B_y^{\rm hll}, \quad B^*_z = B_z^{\rm hll} \,.
\end{equation}
    
\item[(ii)] The normal component of the velocity is recovered from the negative branch of the quadratic equation
\begin{equation} \label{eq::quad_HLLC}
 a\left(v_x^*\right)^2 + bv_x^* + c = 0 \,,
\end{equation}
with coefficients
\begin{equation}
    \begin{array}{lcl}
    a & = & \cF_E^{\rm hll} + \cF_D^{\rm hll} - {\vec B}_\perp^{\rm hll}
            \cdot\cF_{{\vec B}_\perp}^{\rm hll} \,, \\ \noalign{\medskip}
    b & = & -\left(\cF_{m^x}^{\rm hll} + E'^{\rm hll}\right) 
            + |{\vec B}_\perp^{\rm hll}|^2 + |\cF_{{\vec B}_\perp}^{\rm hll}|^2 \,, \\ \noalign{\medskip}
    c & = & m_x^{\rm hll} - {\vec B}_\perp^{\rm hll}
            \cdot\cF_{{\vec B}_\perp}^{\rm hll} \,,         
  \end{array}
\end{equation}
where $E' = E + D$, ${\vec B}_\perp^{\rm hll} = (0,B_y^{\rm hll},B_z^{\rm hll})$ and $\cF_{{\vec B}_\perp}^{\rm hll} = (0,\cF_{B_y}^{\rm hll}, \cF_{B_z}^{\rm hll})$.
    
\item[(iii)] Compute the transverse components of the velocity from
\begin{equation}\label{eq:MB06_vt}
    B_xv_y^* = B_y^*v_x^* - \cF_{B_y}^{\rm hll} \,,\qquad
    B_xv_z^* = B_z^*v_x^* - \cF_{B_z}^{\rm hll}\,.
\end{equation}

Here the $L/R$ subscripts have been removed because of (\ref{eq:MB06_assumption}).

\item [(iv)] Recover the total pressure $p^*$ from
\begin{equation}
    [\cF_E^{\rm hll} + \cF_D^{\rm hll} - B_x^*({\vec v}^*\cdot{\vec B}^*)]v_x^*
     - \left(\DS\frac{B_x^*}{\gamma^*}\right)^2 + p^* - \cF^{\rm hll}_{m^x} = 0 \,,
\end{equation}
where $\vec{v}^* = (v^*_x, v^*_y, v^*_z)$ and $\vec{B}^* = (B_x, B^*_y, B^*_z)$.
    
\item [(v)] Compute the remaining conserved hydrodynamical variables across the contact discontinuity:
\begin{equation}
    \begin{array}{lcl}
    D^* & = & \DS\frac{\lambda - v_x}{\lambda - v_x^*}D \\ \noalign{\medskip}
    E^* & = & \DS\frac{\lambda E - \cF_E + p^*v_x^* - ({\vec v}^*\cdot{\vec B}^*)B_x^*}{\lambda - v_x^*} \\ \noalign{\medskip}
    m_x^* & = & (E'^* + p^*) v_x^* - ({\vec v}^*\cdot{\vec B})B_x^* 
       \\ \noalign{\medskip}
    m_t^* & = & \DS\frac{-B_x^*\left[\left(B_t^*/(\gamma^*)^2\right)
       + ({\vec v}^*\cdot{\vec B}^*)v_t^*\right] + 
       \lambda m_t - \cF_{m_t}}{\lambda - v_x^*}
    \end{array}
    \end{equation}
where $t=y,z$ denotes a generic transverse component and, for the sake of clarity, we have omitted the suffix $(L/R)$.
    
\item [(vi)] Derive the corresponding fluxes from the Rankine-Hugoniot conditions of Eq. (\ref{eq::RH_HLLC}).
    
\end{itemize}

While this approach is fully consistent with the integral average of the solution across the Riemann problem (Eq. \ref{eq::consstateHLLC}), a major drawback is that transverse components of velocity and momentum remain bounded, as $B_x\to0$, only for strictly 2D configurations ($v_z=B_z=0$) while this may not hold in a general 3D vector orientations, as originally noted by \cite{MB2006}.
In these situations (i.e., $\vec{v}^*\cdot\vec{v}^* \ge 1$) we replace the HLLC flux with the the HLL flux (\ref{eq:HLL_flux}).

The limit $B_x = 0$ corresponds to a degenerate situation where slow and Alfv\'en waves propagate at the same speed of the entropy wave.
In this case, not only the density, but also the transverse components of the velocity and magnetic field can experience jumps.
As a consequence, only the normal component of the velocity ($v^*_x$) and the total pressure ($p^*$) are assumed to be continuous.
The previous steps are then modified as follows:
\begin{itemize}
\item [(i)] Find the normal velocity using Eq. (\ref{eq::quad_HLLC}) but with coefficients
\begin{equation}
    \begin{array}{lcl}
    a & = & \cF_E^{\rm hll} + \cF_D^{\rm hll} \\ \noalign{\medskip}
    b & = & -\cF_{m^x}^{\rm hll} + E'^{\rm hll} \\ \noalign{\medskip}
    c & = & m_x^{\rm hll}        
    \end{array}
\end{equation}
where $F^{\rm hll}_E$, $F^{\rm hll}_D$ and $F^{\rm hll}_{m_x}$ are the energy, density and x-momentum component of the HLL flux (\ref{eq:Fhll}).

\item [(ii)] Derive the total pressure from
\begin{equation}
    p^* = \cF_{m_x}^{\rm hll} - (\cF_E^{\rm hll} + \cF_D^{\rm hll})v_x^* \,.
\end{equation}
    
\item [(iii)] Compute the conserved values across the contact discontinuity from
\begin{equation}
    \begin{array}{lcl}
    D^* & = & \DS\frac{\lambda - v_x}{\lambda - v_x^*}D  \,,
    \\ \noalign{\medskip}
    E^* & = & \DS\frac{\lambda E - \cF_E + p^*v_x^*}{\lambda - v_x^*}  \,,
    \\ \noalign{\medskip}
    m_x^* & = & (E'^* + p^*) v_x^*  \,,
    \\ \noalign{\medskip}
    m_t^* & = & \DS\frac{\lambda - v_x}{\lambda - v_x^*}m_t  \,,
    \\ \noalign{\medskip}
    B_t^* & = & \DS\frac{\lambda - v_x}{\lambda - v_x^*}B_t \,,
    \end{array}    
\end{equation}
where, again, $t=y,z$ label a generic transverse component and we have omitted the suffix $(L/R)$ for the clarity of exposition.
    
\item [(iv)] Derive the corresponding fluxes from the Rankine-Hugoniot conditions of Eq. (\ref{eq::RH_HLLC}).
    
\end{itemize}
Notice that, in case of vanishing magnetic field, the latter approach (the one where $B_x = 0$) reduces to the relativistic hydro HLLC solver in \citet{MB2005}.

\subsubsection{Solution of \citet{BalKim2016}}
\label{sec:HLLC_KB}

%

The approach of \cite{KimBal2014} \citep[henceforth HLLC-KB, later corrected in the Appendix B of][] {BalKim2016} presents an improved version of the HLLC solver aimed at resolving the limitations of the previous approach.
For the sake of completeness, we revise here the fundamental steps in order to elucidate some potentially ambiguous aspects in the original formulation.
In particular, Eq. (\ref{eq:MB06_assumption}) is replaced with the weaker requirement
\begin{equation}\label{eq:KB14_assumption}
    \begin{array}{lll}
    v^*_{x,L} = v^*_{x,R}   \qquad & 
    v^*_{y,L} \ne v^*_{y,R} \qquad & 
    v^*_{z,L} \ne v^*_{z,R} \\ \noalign{\medskip}
    p^*_L = p^*_R \qquad & B^*_{y,L} = B^*_{y,R} \qquad & B^*_{z,L} = B^*_{z,R} 
    \end{array}
\end{equation}
that is, the transverse components of velocity are discontinuous across the middle wave while normal velocity, magnetic fields and total pressure are still continuous.

As for the previous HLLC solver, the continuity of $B_y$ and $B_z$ leads to the unique choice
\begin{equation}
    B^*_y = B_y^{\rm hll} \qquad B^*_z = B_z^{\rm hll}.
\end{equation}

By suitable algebraic manipulations, we rewrite the jump condition of the transverse momenta across the outermost waves as 
\begin{equation}\label{eq:KB14_vt}
\begin{array}{ll}
 \Big[ \vec{v}_t^*(m_x - E'\lambda) - p^*\vec{v}_t^*\lambda 
  + \vec{B}_t^*({\vec v}^*\cdot{\vec B}^*)(\lambda - v_x^*) +
    \\ \noalign{\medskip}
  + B_x\DS\frac{\vec{b}_t}{\gamma} - B_x\vec{B}_t^*[1 - (\vec{v}^*)^2] 
   + \vec{m}_t(\lambda - v_x)\Big]_S = 0  \,,
\end{array}
\end{equation}
where, e.g., $\vec{v}_t^*=(0,\, v^*_y,\, v^*_z)$ denotes the transverse velocity vector (the same holds for $\vec{B}^*_t$ and $\vec{b}^*_t$) while, here and in what follows, $S=L$ ($S=R$) implies that the expression applies to the left (right) state.
Eq. (\ref{eq:KB14_vt}) yields indeed a total of 4 equations.

Likewise, it is possible to derive a pair of equations across the left and right waves involving the normal velocity and total pressure:
\begin{equation}
  \begin{array}{l}
   \Big[    (1 - \lambda v_x^*)p^* - B_x^2[1 - (\vec{v}^*)^2] 
  + B_x({\vec v}^*\cdot{\vec B}^*)(\lambda - v_x^*) + \\ \noalign{\medskip}
  +(m_x - \lambda E')v_x^* - m_xv_x
  + B_x  \DS\frac{b_x}{\gamma} + p + \lambda m_x\Big]_S = 0 \,.
  \end{array}
\label{eq::HLLC_KC_vxp}
\end{equation}

Equations (\ref{eq:KB14_vt}) and (\ref{eq::HLLC_KC_vxp}) provide a closed system of 6 equations in the 6 unknown $Q = (\vec{v}_{t,L}^*$,\, $\vec{v}_{t,R}^*$,\, $v_{x}^*$,\, $p^*)$, and, due to its nonlinearity, has to be solved numerically. 
As pointed in \citet{KimBal2014}, the solution of the full set would make the HLLC solution too expensive.
For this reason, the three sets of equations - corresponding, respectively to Eq. (\ref{eq:KB14_vt}) (for the transverse velocities) for $S=L$ and $S=R$, and Eq. (\ref{eq::HLLC_KC_vxp}) for the normal velocity and total pressure - are solved as three $2\times2$ subsystems via multidimensional Newton-Raphson algorithm.
In particular, referring to the left hand sides of Eq. (\ref{eq:KB14_vt}) as, respectively, $G_{y,R}$ and $G_{z,R}$, the corrections to the transverse velocities ($\delta v_{y,R}$, $\delta v_{z,R}$) are recovered as 
\begin{equation}
    \left(\begin{array}{c}
    \delta v_y^* \\ \delta v_z^*
    \end{array}\right)_S   
     = -
    \left(\begin{array}{cc}
    a_{11} & a_{12} \\ a_{21} & a_{22}
    \end{array}\right)^{-1}
    \left(\begin{array}{c}
    G_y \\ G_z
    \end{array}\right)_S \,,
\end{equation}
where the matrix $a$ is the Jacobian matrix, with elements:
\begin{equation}
\begin{array}{l}
    a_{11} = \Big[m_x - \lambda E' - p^*\lambda + (B_y^*)^2(\lambda - v_x^*) 
                      + 2B_xB_y^*v_y^* \Big]_S \,,  \\ \noalign{\medskip}
    a_{12} = \Big[ B_y^*B_z^*(\lambda - v_x^*) + 2B_xB_y^*v_z^* \Big]_S \,,
    \\ \noalign{\medskip}
    a_{21} = \Big[ B_y^*B_z^*(\lambda - v_x^*) + 2B_xB_y^*v_y^* \Big]_S \,,
    \\ \noalign{\medskip}
    a_{22} = \Big[ m_x - \lambda E' - p^*\lambda + (B_z^*)^2(\lambda - v_x^*) 
                     + 2B_xB_z^*v_y^*\Big]_S \,.
\end{array}
\end{equation}
Pressure and normal velocity in this subsystem are kept at the previous iteration level and updated as new values become available during the iteration cycle.

Simultaneously, we solve the $2\times2$ subsystem given by Eqns. (\ref{eq::HLLC_KC_vxp}) for the left and right states.
Denoting with $H_L$ and $H_R$ the left-hand side of Eq. (\ref{eq::HLLC_KC_vxp}), respectively for the left and right state, we get
\begin{equation}
    \left(\begin{array}{c}
    \delta v_{x}^* \\ \delta p^*
    \end{array}\right)   
     = -
    \left(\begin{array}{cc}
    b_{11} & b_{12} \\ b_{21} & b_{22}
    \end{array}\right)^{-1}
    \left(\begin{array}{c}
    H_R \\ H_L
    \end{array}\right)  \,,
\end{equation}
where the elements of the Jacobian matrix are
\begin{equation}
\begin{array}{l}
    b_{11} = \Big[  -\lambda p^* + B_x^2(\lambda + v_x^*) - ({\vec v}^*\cdot{\vec B}^*)B_x + m_x + \lambda E'\Big]_R  \,,
    \\ \noalign{\medskip}
    b_{12} = 1 - \lambda_Rv_x^*   \,,
    \\ \noalign{\medskip}
    b_{21} = \Big[-\lambda p^* + B_x^2(\lambda  + v_x^*) - ({\vec v}^*\cdot{\vec B}^*)B_x + m_x + \lambda E' \Big]_L \,,
    \\ \noalign{\medskip}
    b_{22} = 1 - \lambda_Lv_x^*   \,.
\end{array}
\end{equation}
As for the previous $2\times2$ subsystem, transverse velocity are one iteration late and are taken from Eq. (\ref{eq:KB14_vt}).

Finally, the initial guess to start the Newton-Raphson algorithm is provided by the primitive variables in the HLL state.
The iterative cycle $Q^{*,n+1} = Q^{*,n} + \delta Q$, where $n$ is the iterations number, proceeds until convergence of all the variables is reached (we require an absolute accuracy of $10^{-7}$).

Once the intermediate velocities and total pressure are recovered, the intermediate conserved quantities are computed from
\begin{equation}
    \begin{array}{lcl}
    D^* & = & \DS\frac{\lambda - v_x}{\lambda - v_x^*}D \,, 
    \\ \noalign{\medskip}
    E^* & = & \DS\frac{\lambda E - \cF_E + p^*v_x^* - ({\vec v}^*\cdot{\vec B}^*)B_x^*}{\lambda - v_x^*} \,, 
    \\ \noalign{\medskip}
    {\vec m}^* & = & (E^* + p^* + D^*) {\vec v}^* - ({\vec v}^*\cdot{\vec B}^*){\vec B}^*  \,.
    \end{array}
\end{equation}

The numerical fluxes are then computed from the jump conditions of Eq. (\ref{eq::RH_HLLC}).

When one or more variables fail to converge within 20 iterations, we switch to the simpler HLL method (this has shown, in our experience, to greatly improve the range of applicability of the solver).

We point out, however, that this formulation does not satisfy the state consistency condition given by Eq. (\ref{eq::consstateHLLC}), nor the flux condition (\ref{eq::consfluxHLLC}).
The reason for this incongruity stems from the assumed continuity of $B^*$ across the middle wave while keeping a jump in the transverse velocities. 
As one can immediately verify, in fact, the two assumptions are not compatible with the Rankine-Hugoniot jump conditions for the transverse components of magnetic field across the contact mode, e.g.,
\begin{equation}
  \lambda^* \left(B^*_{y,R} - B^*_{y,L}\right) 
    \ne
     v^*_x \left(B^*_{y,R} - B^*_{y,L}\right) 
   - B_x\left(v^*_{y,R} - v^*_{y,L}\right)  \,,
\end{equation}
which trivially follows from Eq. (\ref{eq:KB14_assumption}) together with the assumption $\lambda^* \equiv v^*_x$.
As a matter of fact, this inconsistency extends also to the momentum and energy jump conditions across the middle wave.

\subsection{HLLD Formulation}
\label{sec:HLLD}
%
The incongruities of the HLLC formulations are fully untangled with the five wave HLLD formulation of \cite{MUB2009} which provides a consistent extension of the original solver by \cite{MiyKus2005} for the MHD equations to the relativistic case.

Here, the Riemann fan is approximated by introducing five waves: two outermost fast shocks, two rotational discontinuities and a contact surface in the middle (slow waves are not considered). 
Since the normal velocity is no longer constant across the rotational waves, the solver is more elaborate than its classical counterpart. 
Still, proper closure is obtained by solving a non-linear scalar equation in the total pressure variable which, for the chosen configuration, has to be constant over the whole Riemann fan.
Hereafter we summarize the procedure and refer the reader to \cite{MUB2009} for the details of the derivation.

The system of jump conditions is written in terms of the 8 unknowns ${D, v_x, v_y, v_z, B_y, B_z, w, p}$ to express states and fluxes:
\begin{equation}
\arraycolsep=1.0pt\def\arraystretch{1.6}
  \begin{array}{l}
  \cU_S = \left(D, w\gamma^2 v_k - b_0b_k, w\gamma^2 - p - b_0b_0, B_k\right)
  \\ \noalign{\medskip}
  \cF_S = \left(Dv_x, w\gamma^2 v_xv_k - b_kb_x + p\delta_{ik},
              w\gamma^2v_x - b_0b_x, B_kv_x - B_xv_k\right)
  \end{array}
\end{equation}
where $S=L, aL, cL, cR, aR, R$ labels one of the possible 6 states (see the third panel in Fig. \ref{fig:riemann_fan}) while $k = x,y,z$ is the subscript for the spatial component.
If $\lambda_S$ separates state $S$ from state $S'$ (clockwise), state and fluxes must satisfy the jump conditions
\begin{equation}\label{eq:hlldR}
  \left(\lambda\cU - \cF\right)_S = \left(\lambda\cU - \cF\right)_{S'}   \,.
\end{equation}

We begin from the states immediately behind the outermost fast waves. 
Dropping the indices $aL$ or $aR$ in the unknowns and using $\lambda$ to denote either $\lambda_L$ or $\lambda_R$, the following expressions for the velocities in the region $aL$ and $aR$ can be derived:
\begin{align}\label{eq:vxa}
 v_x &= \frac{B_x\left(AB_x + \lambda C\right) - 
                \left(A + G\right)\left(p + R_{m_x}\right)}{X},
\\  \label{eq:vta}
 \vec{v}_t &= \frac{Q\vec{R}_{m_t} + \vec{R}_{B_t}\left[C + B^x\left(\lambda R_{m_x} - R_E\right)\right]}{X},
\end{align}
where $\vec{v}^t=(0,v_y, v_z)$, while the different $R_Q$'s denote the components of the array $R = (\lambda\cU - \cF)_{S}$ corresponding to variable Q, with $S=L,R$ for the left or right fast magnetosonic wave, respectively.
The remaining quantities are defined as
\begin{equation}\begin{array}{l}
 A =  R_{m_x} - \lambda R_E + p\left(1-\lambda^2\right), \\
  \noalign{\smallskip}
 G =  \vec{R}_{B_t}\cdot\vec{R}_{B_t},\\
  \noalign{\smallskip}
 C =  \vec{R}_{m_y}\cdot\vec{R}_{B_z} ,\\
  \noalign{\smallskip}
 Q = - A - G + (B^x)^2\left(1-\lambda^2\right),\\
  \noalign{\smallskip}
 X = B_x\left(A\lambda B_x + C\right) - \left(A+G\right)
         \left(\lambda p + R_{E}\right).
\end{array}\end{equation}

Having defined the three components of velocity through the relations
(\ref{eq:vxa})-(\ref{eq:vta}), one immediately obtains the transverse 
magnetic field, total enthalpy, density and energy from the jump conditions across the fast waves:
\begin{equation}
 \vec{B}_t =\frac{\vec{R}_{B_t} - B_x\vec{v}_t}{\lambda - v_x},  \quad
 w   = p + \frac{R_E - \vec{v}\cdot\vec{R}_m}{\lambda - v_x}  \,,
\end{equation}
\begin{equation}\label{eq:hlldD}
 D  = \frac{R_D}{\lambda - v_x},   \quad 
 E  = \frac{R_E + pv_x-(\vec{v}\cdot\vec{B})B_x}{\lambda - v_x}\,,
\end{equation}
while the momentum components follow  from $m_k = (E+p)v_k - (\vec{v}\cdot\vec{B})B_k$

At the Alfv\'en waves, we take advantage of the fact that the expressions
\begin{align}\label{eq:K_simpleL}
 K^k_{cL} = K^k_{aL} & = \left[\frac{R_{m_k} + p\delta_{kx} - R_{B_k}S_x\sqrt{w}}
            {\lambda p + R_E - B_xS_x\sqrt{w}}\right]_L ,
\\ \label{eq:K_simpleR}
 K^k_{cR} = K^k_{aR} & = \left[\frac{R_{m_k} + p\delta_{kx} + R_{B_k}S_x\sqrt{w}}
            {\lambda p + R_E + B_xS_x\sqrt{w}}\right]_R ,
\end{align}
are invariant, respectively, across $\lambda_{aL}$ and $\lambda_{aR}$   \citep{Anile_Pennisi1987} and that $K^x_{aL}=\lambda_{aL}$, $K^x_{aR} = \lambda_{aR}$.
In the previous expressions $S_x={\rm sign}(B_x)$ and the $R$'s are the components of Equation \ref{eq:hlldR} (with $S=L,R$) computed at the outermost waves using either $\lambda = \lambda_L$ or $\lambda=\lambda_R$.

Finally, we impose continuity of the normal velocity across the tangential discontinuity, $v_{x,cL} - v_{x,cR} = 0$, yielding
\begin{equation}\label{eq:hlld_eq}
  \left(K^x_{aR} - K^x_{aL}\right) = B^x\left[
  \frac{1-\vec{K}^2_R}{S_x\sqrt{w_R} - \vec{K}_R\cdot\vec{B}_c}
 +
\frac{1-\vec{K}^2_L}{S_x\sqrt{w_L} + \vec{K}_L\cdot\vec{B}_c}
  \right] \,,
\end{equation}
where $\vec{B}_c=\vec{B}_{cL}=\vec{B}_{cR}$ is the magnetic field in proximity of the contact wave, obtained from the consistency condition between the innermost waves
\begin{equation}\label{eq:Bc}
 \vec{B}_c =
  \frac{\left[\vec{B}(\lambda - v_x) + B_x\vec{v}\right]_{aR}}
       {\lambda_{aR} - \lambda_{aL}} 
         -
  \frac{\left[\vec{B}(\lambda - v_x) + B_x\vec{v}\right]_{aL}}
       {\lambda_{aR} - \lambda_{aL}}  \,.
\end{equation}
Equation (\ref{eq:hlld_eq}) is a nonlinear equation in the total pressure $p$ and has to be solved by means of a standard root-finder method.
Once $p$ has been found with sufficient accuracy, the velocities across the tangential discontinuity can be found by inverting the relation that holds between $K^k$ and the velocity $v_k$. 
The final result is
\begin{equation}\label{eq:vc}
 v_k = K^k - 
 \frac{B_k(1 - \vec{K}^2)}{\pm S_x\sqrt{w} - \vec{K}\cdot{\vec{B}}},
\end{equation}
for $k = x,y,z$. 
Finally, density, energy and momentum are recovered from the jump conditions across $\lambda_{aL}$ and $\lambda_{aR}$ similarly to what done after Equation (\ref{eq:hlldD}).

Once the solution has been found we compute the final interface flux through
\begin{equation}\label{eq:hlld_flux}
  \hat{\cF} = \left\{  \begin{array}{ll}
  \cF_L   
   & \quad {\rm if}\quad 0 < \lambda_L  \\ \noalign{\medskip}
  \cF_{aL} 
   & \quad {\rm if}\quad \lambda_L < 0 < \lambda_{aL}\,, \\ \noalign{\medskip}
    \cF_{aL} + \lambda_{aL} (\cU_{cL} - \cU_{aL}) 
   & \quad {\rm if} \quad \lambda_{aL} < 0 < \lambda_{c}\,, \\ \noalign{\medskip}
    \cF_{aR} + \lambda_{aR} (\cU_{cR} - \cU_{aR}) 
   & \quad {\rm if} \quad \lambda_{c} < 0 < \lambda_{aR}\,,  \\ \noalign{\medskip}
    \cF_{aR} 
   & \quad {\rm if} \quad \lambda_{aR} < 0 < \lambda_{R}\,,  \\ \noalign{\medskip}
    \cF_R    &   \quad {\rm if} \quad \lambda_R < 0 \,, 
  \end{array}  \right. 
\end{equation}
where
\begin{equation}\label{eq:HLLD:Fa}
\begin{array}{lcl}
 \cF_{aL} &=& \cF_L + \lambda_L\left(\cU_{aL} - \cU_L\right)
 \\ \noalign{\medskip}
 \cF_{aR} &=& \cF_R + \lambda_R\left(\cU_{aR} - \cU_R\right)
\end{array}\end{equation}
follow from the jump conditions across the fast waves.
Note that Eq. (\ref{eq:hlld_flux}) corrects the original Eq. [16] reported in \cite{MUB2009} which contains an erroneous speed $\lambda_c$ in the third and fourth cases.

Although Equation (\ref{eq:hlld_eq}) may have, in some circumstances, more than one 
root, the rationale for choosing the physically relevant solution is based on 
positivity of density and on preserving the correct eigenvalue order, i.e., 
$\lambda_{aL} > \lambda_L$, $v_{x,cL} > \lambda_{aL}$ for the left state
and $\lambda_{aR} < \lambda_R$, $v_{x,cR} < \lambda_{aR}$ for the right state.
When one or more of these conditions cannot be met,
we revert to the simpler HLL solver.
\subsection{HLLEM Formulation}
%

The HLLEM Riemann solver has first been proposed by \citet{Einfeldt1988,ERMS1991}, and extended by \citet{DumbserBalsara2016}
to non-conservative hyperbolic systems.
It is an extension of the HLL solution of the Riemann problem \citep{HLL1983}, where the contribution of some selected intermediate waves is included.
The solution of the Riemann problem can be written as three possible states
\begin{equation}
 \cU(0,t) = \left\{
 \begin{array}{lll}
  \cU_L                        & \quad & {\text if}\;\; \lambda_L \geq 0, \\ 
  \cU^{\rm hll} - \cU^{\rm hllem} & \quad & {\text if}\;\; \lambda_L \leq 0 \leq \lambda_R, \\ 
  \cU_R                        & \quad & {\text if}\;\; \lambda_R \leq 0
 \end{array}\right..
\end{equation}
as in the HLL formulation.
The intermediate state, for the sake of clarity, has been split into the HLL component and an antidiffusive term
\begin{equation}
 \cU^{\rm hllem} = \sum_m{\vec R}^m_*(\cU_*)\delta^m_*(\cU_*){\vec L}^m_*(\cU_*)\DS\frac{\lambda_R + \lambda_L}{\lambda_R - \lambda_L}(\cU_R - \cU_L),
\end{equation}
where $m$ indicates the $m-$th intermediate eigenvalue.
The vectors ${\vec R}_*$ and ${\vec L}_*$ are the right and left eigenvectors of the RMHD equations, where the subscript $*$ means 
that they are computed from the average of the conserved variables, while the matrix $\delta_*$ is computed as follows:
\begin{equation}
  \delta_*^m(\cU) = 1 - \frac{\lambda_{m,*} - |\lambda_{m,*}|}{2\lambda_L}- \frac{\lambda_{m,*} + |\lambda_{m,*}|}{2\lambda_R}.
\end{equation}
The corresponding numerical fluxes are:
\begin{equation}
\hat{\cF} = \left\{
\begin{array}{lll}
  \cF_L                        & \quad & {\text if}\;\; \lambda_L \geq 0, \\
  \cF^{\rm hll} - \cF^{\rm hllem} & \quad & {\text if}\;\; \lambda_L \leq 0 \leq \lambda_R, \\
  \cF_R                        & \quad & {\text if}\;\; \lambda_R \leq 0, 
\end{array}
\right.,
\end{equation}
where $\cF^{\rm HLLEM}$ is the antidiffusive term
\begin{equation}
 \cF^{\rm hllem} = \DS\left(\frac{\lambda_R\lambda_L}{\lambda_R - \lambda_L}\right)\sum_m\delta^m_*{\vec R}^m_*\left[{\vec L}^m_*\cdot(\cU_R - \cU_L)\right]
\end{equation}
Clearly, such solver becomes complete if all of the intermediate waves are considered, although, as pointed by \citet{BalKim2016,PunBal2016}, the eigenvectors for the fast and slow magnetosonic waves are very expensive to evaluate computationally.
Therefore we consider, as in \citet{PunBal2016}, the 5-wave HLLEM formulation, which captures only contact discontinuities and Alfv\'en waves, with eigenvalues, respectively,
\begin{equation}
 \lambda_e = v^x \,, \qquad
 \lambda_{a,\pm} = \DS\frac{b^x \pm\sqrt{w_T}u^i}{b^0 \pm\sqrt {w_T}\gamma}\,,
\end{equation}
where $w_T = \rho h + b^2$ is the total enthalpy.

Finally we note that in this paper we provide a slightly modified strategy from \citet{Anton_etal2010} to recover the left and right eigenvectors corresponding to the contact and Alfv\'en waves.
This shown in detail in Appendix \ref{sec:Appeigenvec}.

\subsection{GFORCE Formulation}
%

The generalized FORCE flux \citep{TorTit2006} is a generalization of the First ORder CEntred (FORCE) scheme and it consists of a convex average of the Lax-Wendroff ($\cF^{LW}$) and Lax-Friedrichs ($\cF^{LF}$) fluxes:
\begin{equation}
    \cF = \omega_g\cF^{LW} + (1-\omega_g)\cF^{LF} \,.
\end{equation}
where $\omega_g \in[0,1]$.
Here the Lax Wendroff flux is computed as $\cF^{LW} = \cF(\cU^{LW})$, where $\cF$ is given by Eq.  (\ref{eq::flux}), and 
\begin{equation}
    \cU^{LW} = \frac{\cU_R + \cU_L}{2} - \frac{\tau}{2}(\cF_R - \cF_L) \,,
\end{equation}
while the Lax-Friedrichs flux is defined by 
\begin{equation}
   \cF^{LF} = \frac{\cF_R + \cF_L}{2} - \frac{1}{2\tau}(\cU_R - \cU_L) \,.
\end{equation}
In the original formulation by \citet{TorTit2006}, the variable $\tau$ (which has the dimensions of inverse velocity) is set to be $\tau  = \Delta t/\Delta x$.
However, we choose to follow the formulation of \citet{MigZan2021}, where $\tau$ is the inverse of the local maximum signal velocity:

\begin{equation}
 \tau = [\max(|\lambda_L|,|\lambda_R|)]^{-1}.
\end{equation}

The parameter $\omega_g$ can be tuned according to stability and monotonicity criteria, as thoroughly explained in \citet{TorTit2006, Toro_Book2009}.
While $\omega_g=0$ reduces the scheme to the simple Lax-Fridrichs solver, the choice $\omega_g=1/2$ yields the FORCE flux which is precisely the arithmetic mean between the Lax-Friedrichs and Lax-Wendroff fluxes. 
This scheme has reduced dissipation when compared to the LF solver and it corresponds to a monotone scheme with the maximum region of monotonicity, without resorting to wave propagation information.
Larger values of $\omega_g$ are also possible without violating the monotonicity region by choosing 
\begin{equation}\label{eq:GFORCE_omega}
    \omega_g = \frac{1}{1 + c_g},
\end{equation}
where $c_g\in[0,1]$ is the Courant number.
Eq. (\ref{eq:GFORCE_omega}) will be used by default unless otherwise stated.

%% file: results.tex
\section{Numerical Benchmarks}
\label{sec::Tests}
%
%
%

\begin{table*}
\caption{Initial conditions for left and right states (column 2-9), adiabatic index (col 10), final time (col. 11) and number of cells (col 12) for the 1D test problems. 
Here \quotes{CW} and \quotes{RW} refer to the isolated contact and rotational wave, 
while \quotes{ST1}-\quotes{ST4} corresponds to the different shock tubes.
\label{tab:1Dtests} }
 \centering
\begin{tabular}{rccccccccccc}
 \hline
  Case & $\rho$ & $p$ & $v_x$ & $v_y$ & $v_z$ & $B_x$ & $B_y$ & $B_z$ & $\Gamma_{\rm eos}$ & $t_f$ & $N_x$ \\
 \hline 
  CW L & 10.0 & 1.0 & 0.0 & 0.7 & 0.2 & 5.0 & 1.0 & 0.5 & \multirow{2}{*}{5/3} &       
     \multirow{2}{*}{1} & \multirow{2}{*}{40} \\
     R & 1.0  & 1.0 & 0.0 & 0.7 & 0.2 & 5.0 & 1.0 & 0.5 \\
     \hline
  RW L & 1.0 & 1.0 & 0.4 & -0.3 & 0.5 & 2.4 & 1.0 & -1.6 & \multirow{2}{*}{5/3} &       
     \multirow{2}{*}{1} & \multirow{2}{*}{40} \\
     R & 1.0 & 1.0 & 0.377237 & -0.482389 & 0.424190 & 2.4 & -0.1 & -2.178213 \\
     \hline
 ST1 L & 1.0 & 1.0 & 0.0 & 0.0 & 0.0 & 0.5 & 1.0 & 0.0 & \multirow{2}{*}{2.0} &       
     \multirow{2}{*}{0.4} & \multirow{2}{*}{400} \\
     R & 0.125  & 0.1 & 0.0 & 0.0 & 0.0 & 0.5 &  -1.0 & 0.0 \\
     \hline
 ST2 L & 1.08 & 0.95 & 0.4 & 0.3 & 0.2 & 2.0 & 0.3 & 0.3 & \multirow{2}{*}{5/3} &       
     \multirow{2}{*}{0.55} & \multirow{2}{*}{800} \\
     R & 1.0 & 1.0 & -0.45 & -0.2 & 0.2 & 2.0 & -0.7 & 0.5 \\
     \hline
 ST3 L & 1.0 & 0.1 & 0.999 & 0.0 & 0.0 & 10.0 & 7.0 & 7.0 & \multirow{2}{*}{5/3} &       
     \multirow{2}{*}{0.4} & \multirow{2}{*}{400} \\
     R & 1.0  & 0.1 & -0.999 & 0.0 & 0.0 & 10.0 &  -7.0 & 7.0 \\
     \hline
 ST4 L & 1.0 & 5.0 & 0.0 & 0.3 & 0.4 & 1.0 & 6.0 & 2.0 & \multirow{2}{*}{5/3} &       
     \multirow{2}{*}{0.55} & \multirow{2}{*}{800} \\
     R & 0.9 & 5.3 & 0.0 & 0.0 & 0.0 & 1.0 & 5.0 & 2.0 \\
     \hline
\end{tabular}
\end{table*}

We now compare, in terms of accuracy and computational efficiency, the designated Riemann solvers through a set of one- and multi-dimensional leading benchmark solutions commonly employed.
All computations are carried out using the PLUTO code for plasma astrophysics \citep{Mignone_PLUTO2007} where algorithms are readily available.
One-dimensional tests are performed using a $1^{\rm st}$-order scheme with flat reconstruction and explicit Euler time stepping.
In two- or three-dimensions we employ $2^{\rm nd}$-order time integration using a Runge-Kutta algorithm \citep{GST2001} and linear reconstruction with slope limiters. 
The divergence-free constraint of magnetic field is controlled using the constrained transport method with the CT-Contact scheme by \cite{GS2005} to compute the electromotive force at zone edges.
Unless otherwise stated, we set the CFL number  to $0.8$, $0.4$ and $0.25$, respectively, for 1D, 2D and 3D computations.

\subsection{Isolated Contact and Rotational Waves}
%

\begin{figure}
\centering
\includegraphics[width=0.49\textwidth]{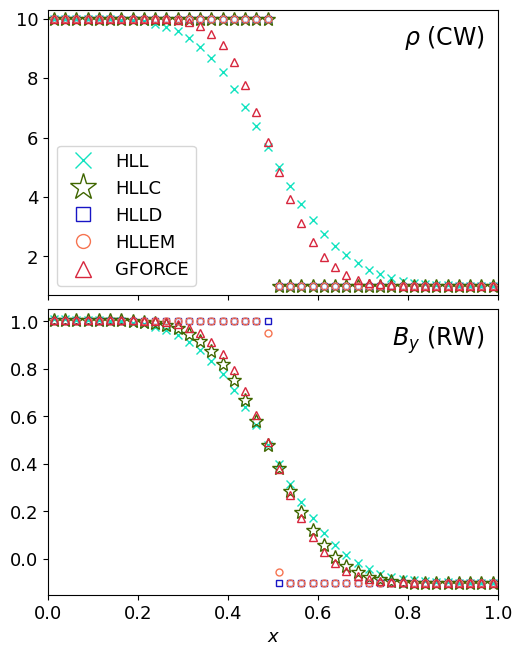}
\footnotesize
\caption{Top panel: density profile at $t=1$ for a single contact wave. 
Bottom panel: $y$-component of magnetic field at $t=1$ in the case of an isolated rotational discontinuity. 
Different solvers are labeled in the legend.}
\normalsize
\label{fig:CW+RW}
\end{figure}

We begin our benchmark section by testing the solvers ability in capturing isolated contact and rotational waves, as shown in \cite{MUB2009}.
The initial conditions together with the final time and number of points are listed in the $1^{\rm st}$ and $2^{\rm nd}$ row in Table \ref{tab:1Dtests}.

In the case of an isolated contact wave, the top panel in Fig. \ref{fig:CW+RW} shows that the numerical solutions recovered HLLD, HLLC and HLLEM solvers resolve the contact mode exactly, while HLL and GFORCE spread the discontinuity over several zones, although the latter performs noticeably better than the former ($\sim 22$ vs. $\sim 16$ cells, respectively).

For a single rotational wave, only the HLLEM and HLLD solver catch the correct behavior as can be inferred from the bottom panel of Fig. \ref{fig:CW+RW} showing the profile of $B_y$.
On the contrary, results obtained with the other solvers (i.e. HLL, HLLC and GFORCE) present significant amount of numerical diffusion by spreading the initial jump over $\sim 10$ computational zones.
No difference has been found between HLLC-MB and HLLC-KB.

\subsection{Shock Tubes}
%

Next, we consider a set of shock-tube problems, by following the same standards adopted by \cite{MUB2009,Anton_etal2010}.
In order to strengthen and compare the influence of different Riemann solvers on the solution, a flat ($1^{\rm st}$-order) reconstruction is used for all of them.  
Numerical results are compared with the exact numerical solution available from  \citet{GR2006} by computing discrete errors in $L1$-norm.

Here we employ HLLC-MB only since HLLC-KB gives essentially the same results.
For completeness, we list the initial conditions in Table \ref{tab:1Dtests}.

\subsubsection{Shock Tube 1 (ST1)}
%

\begin{figure*}
\centering
\includegraphics[width=0.99\textwidth]{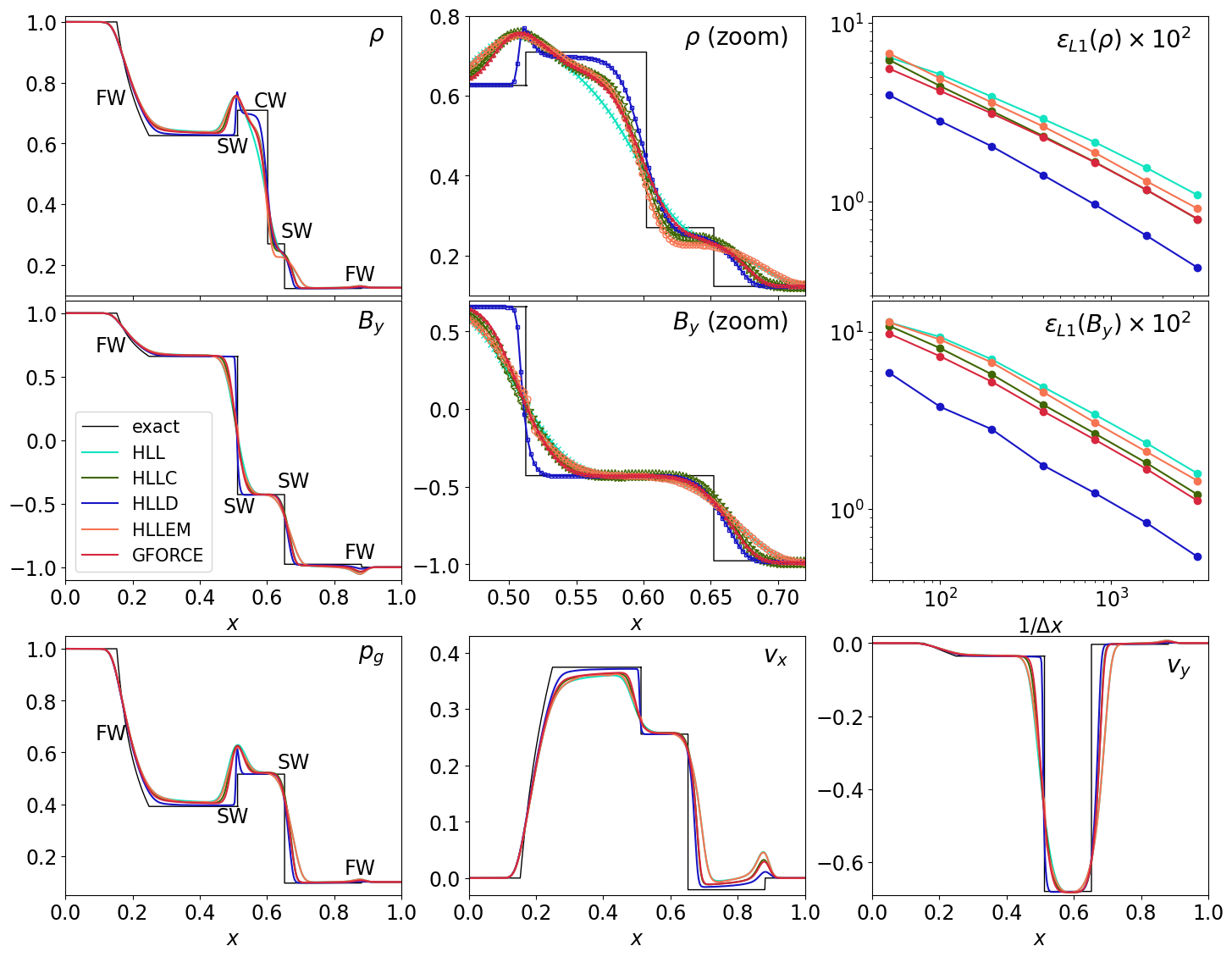}
\footnotesize
\caption{Numerical results for the $1^{\rm st}$ shock tube (ST1) at $t=0.4$ with $400$ grid zones and different Riemann solvers.
Top panels (left to right): density profile, closeup view across the contact mode and $L1$-norm errors. 
Middle panels: same as before but for the $y$-component of magnetic field.
Bottom panels: gas pressure, $x$- and $y$-components of velocity.
}
\normalsize
\label{fig:ST1}
\end{figure*}

This test, performed previously by \cite{Balsara2001,MB2006, MUB2009,Anton_etal2010}, contains only co-planar vectors on either side of the discontinuity and thus no rotational wave can form in the solution. 
The approximate structure of the Riemann fan is shown in Fig. \ref{fig:ST1} at $t = 0.4$ for various solvers.
HLLD performs the best, by showing enhanced resolution and better accuracy in proximity of all waves: at the fast rarefaction tail (FW, $x\sim 0.25$), the compound wave (SW, $x\approx 0.5$), the contact mode (CW, $x\approx 0.6$), the right-facing slow shock (SW, $x\approx 0.65$) and the fast shock (FW, $x\approx 0.9$).
A zoomed view across the contact wave (top central panel), reveals that also HLLC, HLLEM and GFORCE capture equally well this mode while relatively poor resolution is observed at the slow shock (central panel, closeup view on $B_y$), where the HLLEM solver shows a slightly worse performance than the HLLC and GFORCE Riemann solvers.
This is also confirmed from the $L1$-norm error of density and $y$-component of magnetic field (rightmost top and middle panels), indicating that the HLLD has considerable smaller errors, followed by GFORCE, HLLC and, close-by, by HLLEM and HLL.

This result should not be surprising, since the characteristic information restored in the HLLEM solver is based on a linearization process and can cope specifically only with those waves it was initially intended to resolve (contact and rotational waves in our implementation).
On the contrary, HLLC and HLLD solvers stem from a nonlinear approximation to the Riemann fan, in conformity with the integral representation of the Riemann fan where, for mathematically consistency, fewer conditions are imposed on the internal wave structure.
This leads to a set of jump conditions where flow variables can experience jumps not necessarily corresponding to the specific wave (e.g. contact or Alfv\'en) they were originally designed for.

\subsubsection{Shock Tube 2 (ST2)}
%

\begin{figure*}
\centering
\includegraphics[width=0.99\textwidth]{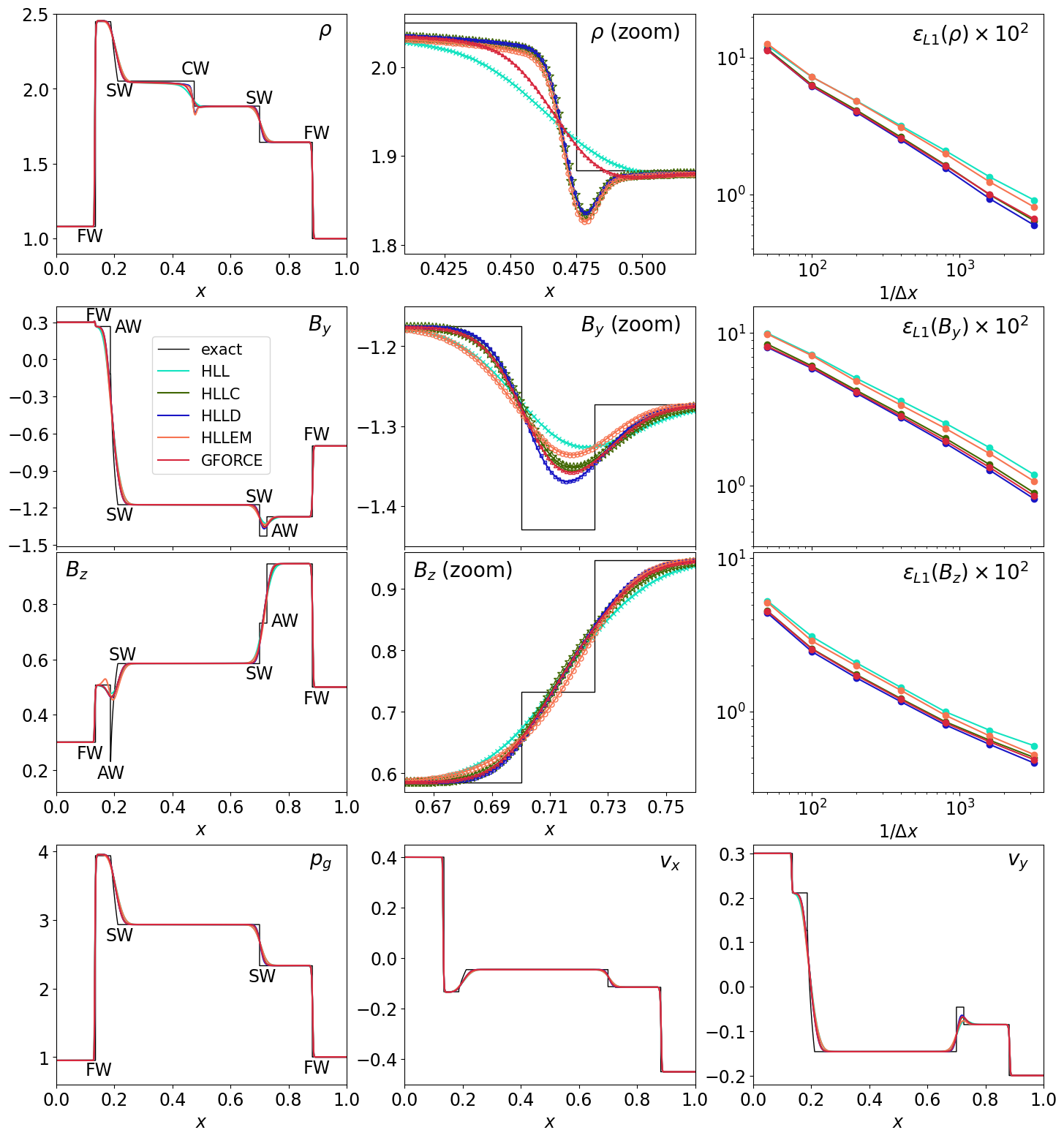}
\footnotesize
\caption{Results for the $2^{\rm nd}$ shock tube (ST2) at $t=0.55$ using a $1^{\rm st}$-order scheme with $800$ grid zones.
Top row (left to right): density profile, closeup view across the contact wave and $L1$-norm errors. 
Second (third) row: $y$- ($z$-) component of magnetic field, closeup view across the slow and Alfv\'en wave and $L1$-norm errors.
Bottom row: profiles for gas pressure, $x$- and $y$-velocity components.
}
\normalsize
\label{fig:ST2}
\end{figure*}

This test, also considered in \cite{Balsara2001, MUB2009,Anton_etal2010,BeckStone2011}, features a non-planar Riemann problem leading to a change in orientation of the transverse magnetic field across the Riemann fan.
The emerging wave pattern consists of a contact wave (CW at $x \approx  0.475$) separating a left-going fast shock (FW, $x \approx 0.13$), Alfv\'en wave (AW, $x\approx0.185$) and slow rarefaction (SW, $x\approx 0.19$) from a slow shock (SW, $x\approx 0.7$), Alfv\'en wave (AW, $x\approx0.725$) and fast shock (FW, $x\approx 0.88$) heading to the right.

Results, at $t=0.55$ are shown in Fig. \ref{fig:ST2}.
Now the differences between the chosen Riemann solvers are less pronounced.
Such alikeness is reflected in the $L1$-norm errors in the right panels, where HLLD, GFORCE and HLLC show similar accuracy, while the HLLEM and HLL solvers exhibit somewhat larger errors.

The contact mode is well resolved by all solvers (although with spurious undershoots, see the top central panel), exception made for HLL and GFORCE which are not designed to minimize the diffusion across the contact wave.
As for the previous test, we again note that GFORCE spreads the contact wave over fewer zones when compared to HLL.

The slow modes, which are not designed to be resolved by any of such solvers (see the central panels of the 2$^{\rm nd}$ and $3^{\rm rd}$ rows), are better captured by HLLD, GFORCE and HLLC, while the HLLEM and the HLL solvers behave in the same way.
Since the slow and the Alfv\'en modes are very close to each other, the accuracy of the HLLEM solver results strongly reduced despite its ability to capture the rotational discontinuities.
Furthermore, the HLLEM shows a non-physical overshoot behind the left Alfv\'en wave (see left panel of the 3-rd row), which vanishes at higher resolution.

The previous considerations are verified more quantitatively by the three error plots in the rightmost panels, again confirming that the HLLD Riemann solver yields the most accurate results followed, in decreasing order of accuracy, by GFORCE, HLLC, HLLEM and HLL.
Note that, as in the previous test, while the GFORCE and the HLLC solvers have the same level of accuracy in the density, the GFORCE performs slightly better when looking at other variables because of the reduce dissipation along the slow modes.

\subsubsection{Shock Tube 3 (ST3)}
%

\begin{figure*}
\centering
\includegraphics[width=0.99\textwidth]{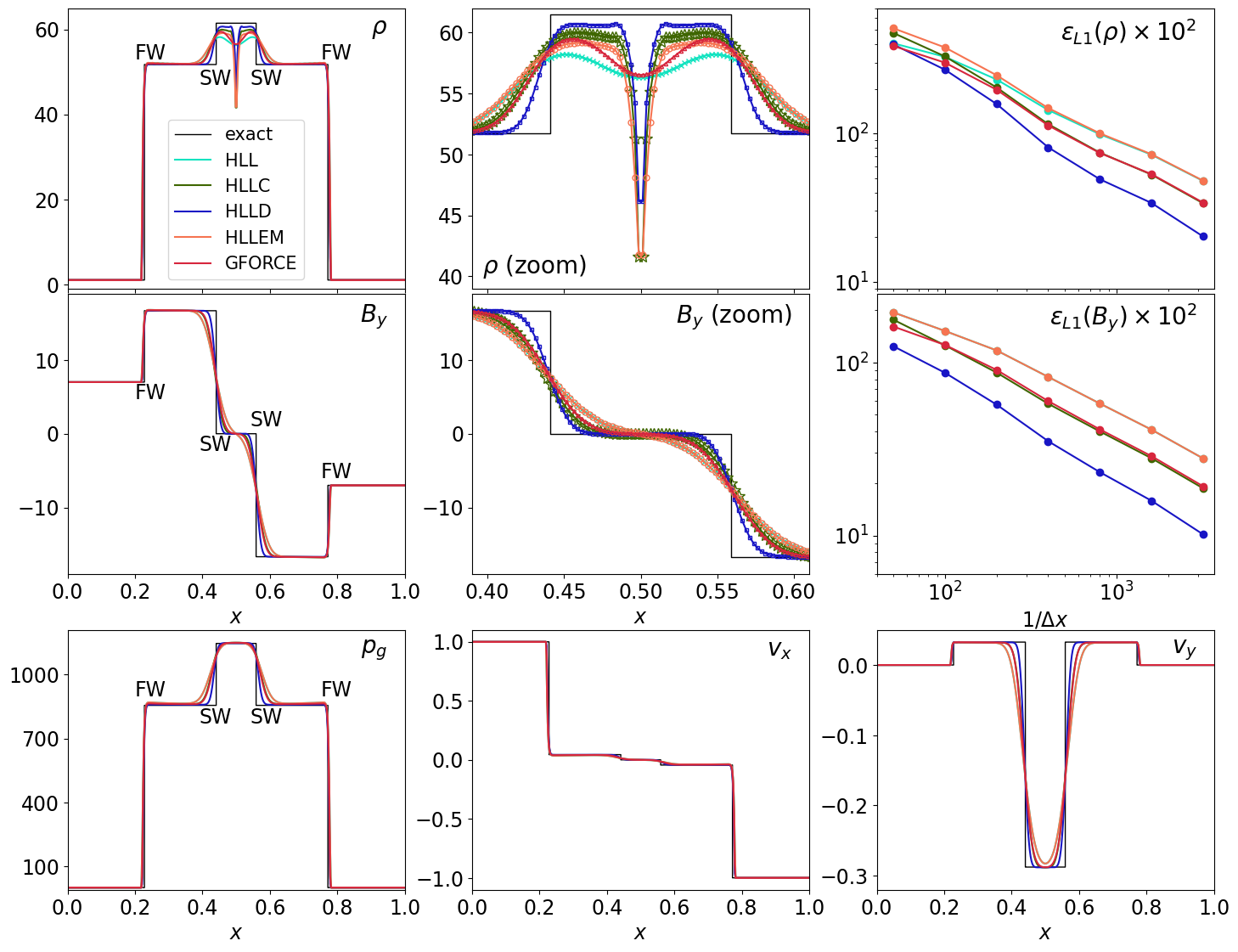}
\footnotesize
\caption{Same as Fig. \ref{fig:ST1} but for the $3^{\rm rd}$ shock tube (ST3).
}
\normalsize
\label{fig:ST3}
\end{figure*}

The initial conditions for this test problem, given in Table \ref{tab:1Dtests}, sets the stage for two oppositely colliding relativistic streams.
This test problem has been previously considered also by \cite{Balsara2001,MB2006, MUB2009,Anton_etal2010}.

The impact generates two strong relativistic fast shocks ($x\approx 0.23$, $x\approx0.77$) heading outwards in opposite directions about the impact point at $x = 0.5$, see Fig. \ref{fig:ST3}. 
Behind, two slow shocks (SW, $x\approx 0.44$ and $x\approx 0.56$) delimit a high-pressure and constant density region. 
Similarly to ST1, this is a co-planar problem and no rotational mode can develop in the solution.
No contact wave is formed either.
Because of the absence of contact and Alfv\'en waves, the HLLEM and the HLL solvers are not distinguishable in every variable (but the density).

We notice that the GFORCE solver suffers from negative values of gas density and pressure caused by the strong gradients of the fast shocks.
In order to overcome such issue, we switched to the FORCE flux ($\omega_g = 1/2$) which still yields reduced numerical diffusion when compared to the HLL solver.

The spurious density dip at the initial collision point ($x =0.5$) is a symptom of the \quotes{wall heating} phenomenon  occurs \citep{Noh1987,DM1996}.
Because of the larger numerical diffusion, the HLL and the FORCE solvers are less prone to such pathology (the error respect to the analytical solution at $x = 0.5$ is, respectively, $\sim8.4\%$ and $\sim7.9\%$).
On the other hand, HLLD, HLLC and HLLEM feature a deeper \quotes{hole} in the rest-mass density (the numerical undershoot is, respectively, $\sim25\%$, $\sim32.3\%$ and $\sim32\%$).
As a consequence, as shown in the top right panel, the HLL and the HLLD solvers shows a similar accuracy in the density at low resolution.
As the number of grid cells increases, this density undershoot is progressively confined to a smaller fraction of the computation domain, leading to a better accuracy in the HLLD solver.
This feature is not found in other variables, where the HLLD solver performs significantly better than the other solvers.
From the error plots, we evince that the FORCE solver performs better than the HLLEM and HLL solver with errors comparable to the HLLC-MB.

\subsubsection{Shock Tube 4}
%

\begin{figure*}
\centering
\includegraphics[width=0.99\textwidth]{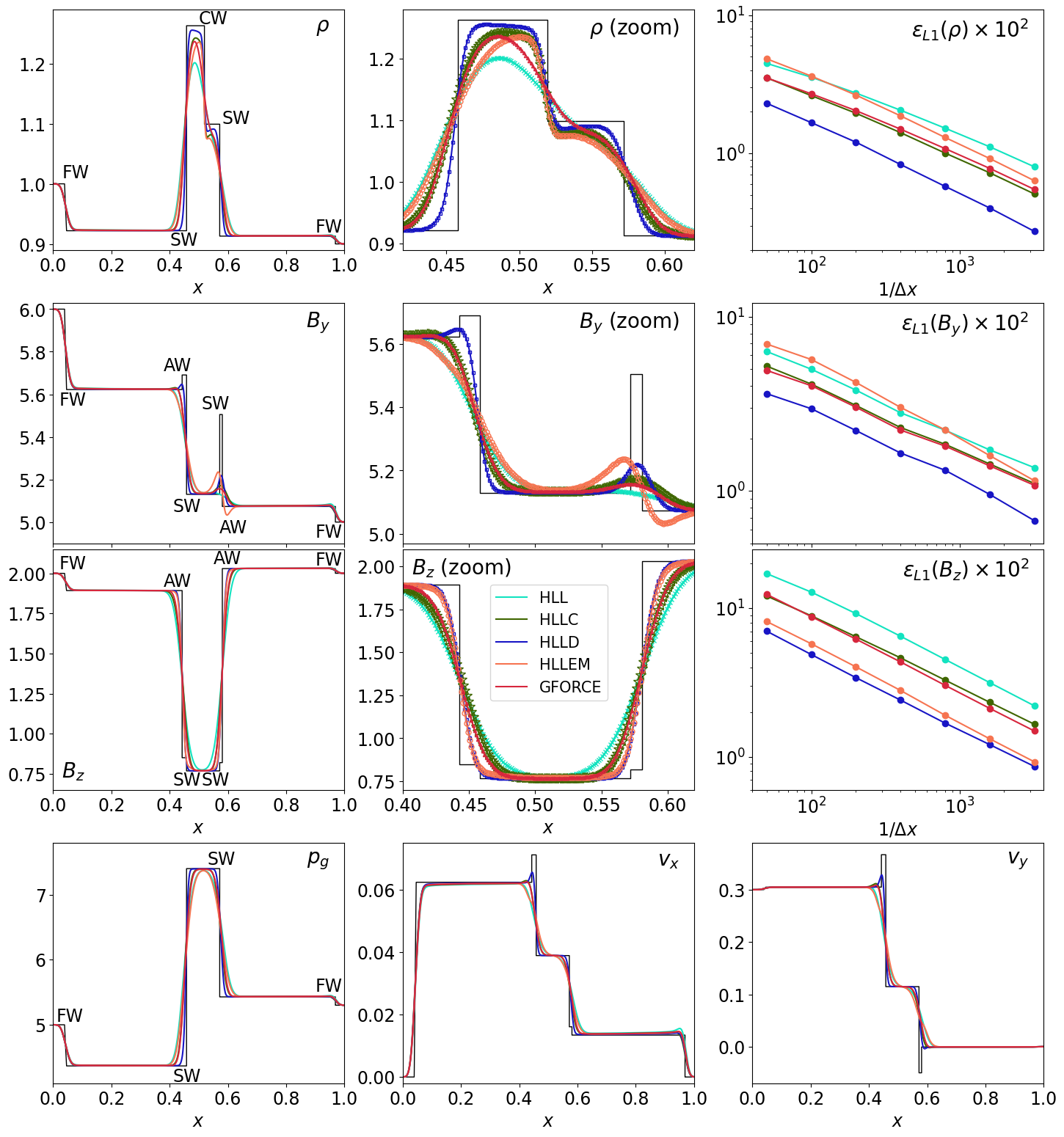}
\caption{Same of \ref{fig:ST2} but for the $4^{\rm th}$ shock tube (ST4).
}
\label{fig:ST4}
\end{figure*}

The initial discontinuity of ST4, which corresponds to the \quotes{Generic Alfv\'en test} of \citet{GR2006}, leads to solution consisting of 7 waves: a fast rarefaction ($x\approx0.04$), a rotational wave ($x\approx0.44$), a left-going slow shock ($x\approx0.46$), a contact discontinuity ($x\approx0.52$), a right-going slow shock ($x\approx0.57$), a rotational wave ($x\approx0.58$) and a fast shock ($x\approx0.97$).

Results, plotted in Fig. \ref{fig:ST4}, demonstrate that the HLLD is able to reach better accuracy than all the other solvers, as in the previous tests.
Looking at the left-going slow and rotational modes ($B_y$ profile in the central column, $2^{\rm nd}$ row), we observe that the HLLD solver is the only one able to resolve both  modes, while all the other solvers are unable to capture them.

Again, we remark that HLLC and GFORCE solvers give comparable results.
In particular, the HLLC solver provides a better resolution only at the contact wave (giving better results for the density error), while fast, slow and Alfv\'en modes are resolved with comparable accuracy.

Since the Alfv\'en mode is very prominent in the $z-$components of the magnetic field (central panel, $3^{\rm rd}$ row), the HLLEM is able to reach a higher precision, comparable to the  HLLD solver. 
In addition, the HLLEM and the HLLD solvers are the only able to capture the right-going slow and Alfv\'en modes (visible from the $y$-component of $\vec{B}$, central panel, $2^{\rm nd}$ row).
Still, the HLLEM solver presents some unphysical undershoots in the $y$-component of the magnetic field, which (contrary to the expectation) severely affects the error. 
Such issue lowers at larger resolutions (see the error plots in the right panel, $2^{\rm nd}$ column).
The other solvers show some barely visible structure (HLLC and GFORCE) or just a single blended wave (HLL solver).

\subsection{Circularly Polarized Alfv\'en Waves}
%

\begin{figure*}
\centering
  \includegraphics[width=0.99\textwidth]{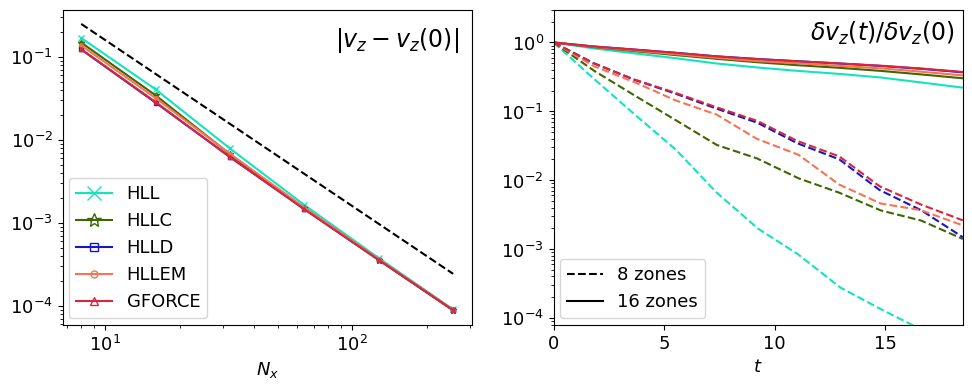}
  \caption{Left panel: $L_1$-norm error for $v_z$ in the circularly polarized 
   Alfv\'en test problem after one period $T = 1/\sqrt{2}v_A \approx 1.851$
   and different Riemann solvers (see the legend).
   Right panel: amplitude decay as a function of time using, respectively, $8$ 
  (dashed lines) and  $16$ (solid lines) zones per wavelength.
  }
  \label{fig:CPA}
\end{figure*}

Next, we consider the propagation of large amplitude, circularly polarized (CP) Alfv\'en waves on a two-dimensional unit square domain, as in  \cite{dZ_etal2007}.
The initial condition consists of a region of uniform density and pressure ($\rho=p=1$) while magnetic field and velocity, for a wave front propagating along the $x'$ direction, are given by
\begin{equation}\label{eq:CPA1}
 \vec{B}' =   B_0\left(1,\, \eta\cos\phi,\, \eta\sin\phi\right), \quad
 \vec{v}' =  -v_A\left(0,\, \frac{B'_y}{B_0},\, \frac{B'_z}{B_0}\right)
\end{equation}  
where $\phi = k'x'$ is the wave phase, $k'$ is the wavenumber.
In Eq. (\ref{eq:CPA1}) $B_0=1$ is the (constant) magnetic field component in the direction of propagation, $\eta=1$ is the amplitude and the Alfv\'en velocity $v_A$ is computed from 
\begin{equation}\label{eq:CPA2}
  v_A^2 = \frac{2\alpha}{1 + \sqrt{1 - 4\eta^2\alpha^2}},\quad{\rm where}\quad
 \alpha = \frac{B_0^2}{w_g + B_0^2(1 + \eta^2)}
\end{equation}
and $w_g = \rho + \Gamma p /(\Gamma -1)$.
This yields $v_A\approx 0.382$ for our parameter choice (we use $\Gamma=4/3$).
The previous conditions provide an exact wave solution of the RMHD equations provided $\phi\to \phi-\omega t$, where $\omega = k'v_A$ is the angular frequency and are thus valid for arbitrary amplitude $\eta$.

We perform the test on a 2D Cartesian domain $x\in[0,L_x]$, $y\in[0,L_y]$ with $L_x = L_y = 1$ and rotate the coordinate system by an angle $\alpha$ around the $z$-axis, so that vectors are rotated according to 
\begin{equation}
  \vec{q} = \tens{R}\,\vec{q}' \,,\quad \quad{\rm with} \qquad
  \tens{R} = \left(\begin{array}{ccc}
              \cos\alpha  & -\sin\alpha   &  0 \\ \noalign{\medskip}
              \sin\alpha  &  \cos\alpha   &  0 \\ \noalign{\medskip}
                     0  &           0   &  1 \\ \noalign{\medskip}
          \end{array}\right)\,, 
\end{equation}
where $\vec{q}$ is a generic vector in the rotated (computational) frame while $\vec{q}'$ is the corresponding vector in the 1D (unrotated) frame.
The wave vector components are chosen so that exactly one wavelength fits along the domain sizes, $\vec{k}= (2\pi/L_x)\hvec{e}_x + (2\pi/L_y)\hvec{e}_y$ (note that $\phi$ is invariant under rotations).
Computations are performed with $N_x\times N_x$ grid zones using a Courant number $C_a=0.4$.

In the left panel of Fig. \ref{fig:CPA} we measure, as a function of the resolution $N_x$, the accuracy of the selected Riemann solvers by computing, after one period $T=1/(\sqrt{2}v_A)$, the $L_1$ norm errors of the vertical component of velocity $v_z$.
Second-order accuracy is obtained with all solvers, although HLL has a slightly large errors at small resolutions.

In the right panel of Fig. \ref{fig:CPA} we compare the dissipative properties of the different solvers by measuring the decay of the wave amplitude, defined as $\delta v^z=\max(v^z) - \min(v^z)$ (normalized to its initial value) up to ten revolutions, using $8$ and $16$ zones per wavelength, respectively.
Overall, the HLLD and GFORCE Riemann solvers yield the smallest dissipation, followed by HLLEM and HLLC and lastly by HLL.
At low resolution, the wave amplitudes decrease approximately by $\sim 10^{-3}$ of the nominal value (for the first four solvers) while to $\sim 10^{-5}$ for the HLL solver.
By increasing the resolution to $16$ zones, differences are less pronounced and wave amplitudes drop to $\sim 0.37$ (HLLD and GFORCE), $\sim 0.33$ (HLLEM), $\sim 0.30$ (HLLC) and $\sim 0.22$ (HLL).

We point out that the smoothness of the solution allowed the GFORCE scheme to be run with $\omega$ defined as in Eq. (\ref{eq:GFORCE_omega}) with $c_g = 0.4$.
Smaller values of $\omega$ (higher values of $c_g$) bias the scheme towards a more diffusive behavior.
In the limit $\omega\to 1/2$ one retrieves the FORCE scheme which yields results comparable with the HLLC solver for this problem.

\subsection{Blast Waves (2D, 3D)}
%

\begin{figure*}
\centering
\includegraphics[width=0.99\textwidth]{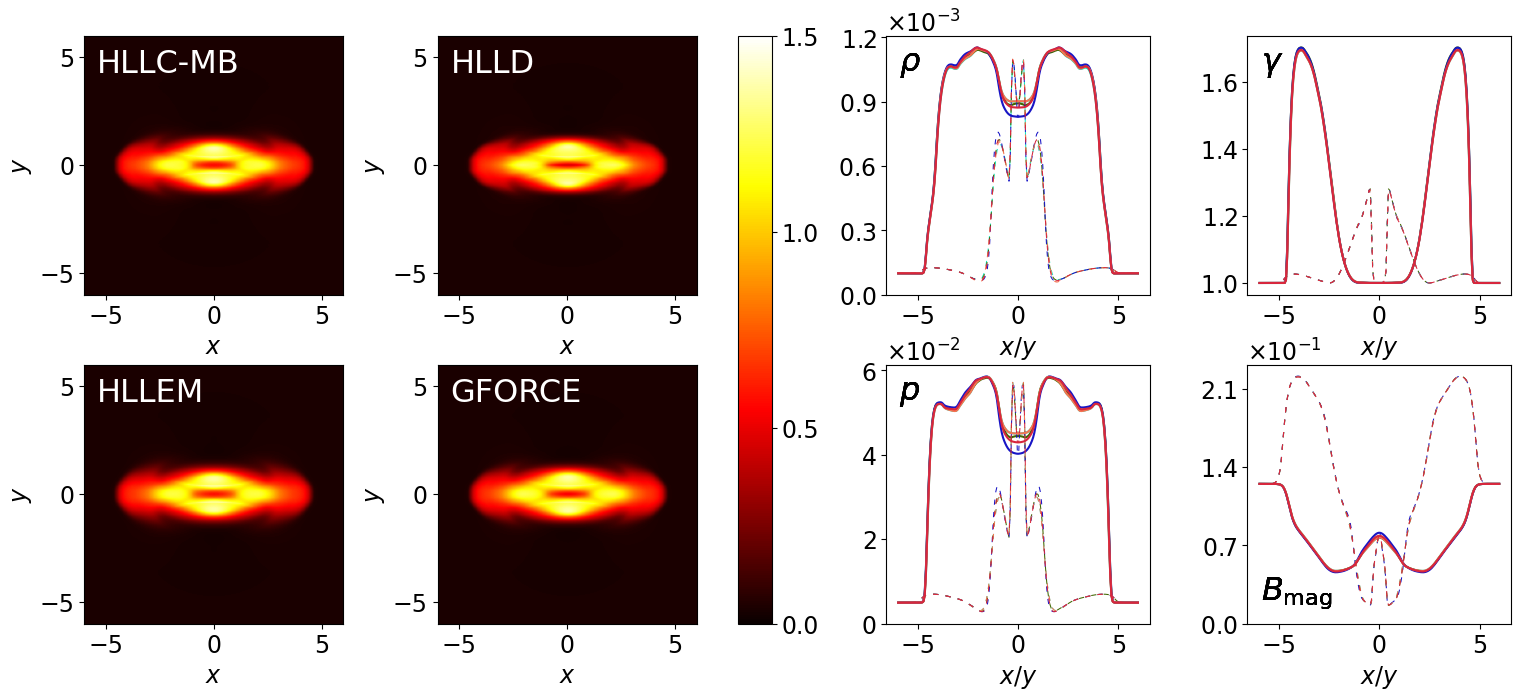}
\footnotesize
\caption{Results for the 2D blast wave problem at $t=4$, for $B_0=0.5$ and $\phi=0^\circ$. In the left half we show Colored maps of the plasma $\beta=2p/B^2$ (left) for different Riemann solvers while in the right half we present 1D profiles along the $x$-axis (solid line) and $y$-axis (dotted lines).
Color convention is the same adopted for previous tests.}
\normalsize
\label{fig:BW2D_B0.5_phi0}
\end{figure*}

\begin{figure*}
\centering
\includegraphics[width=0.99\textwidth]{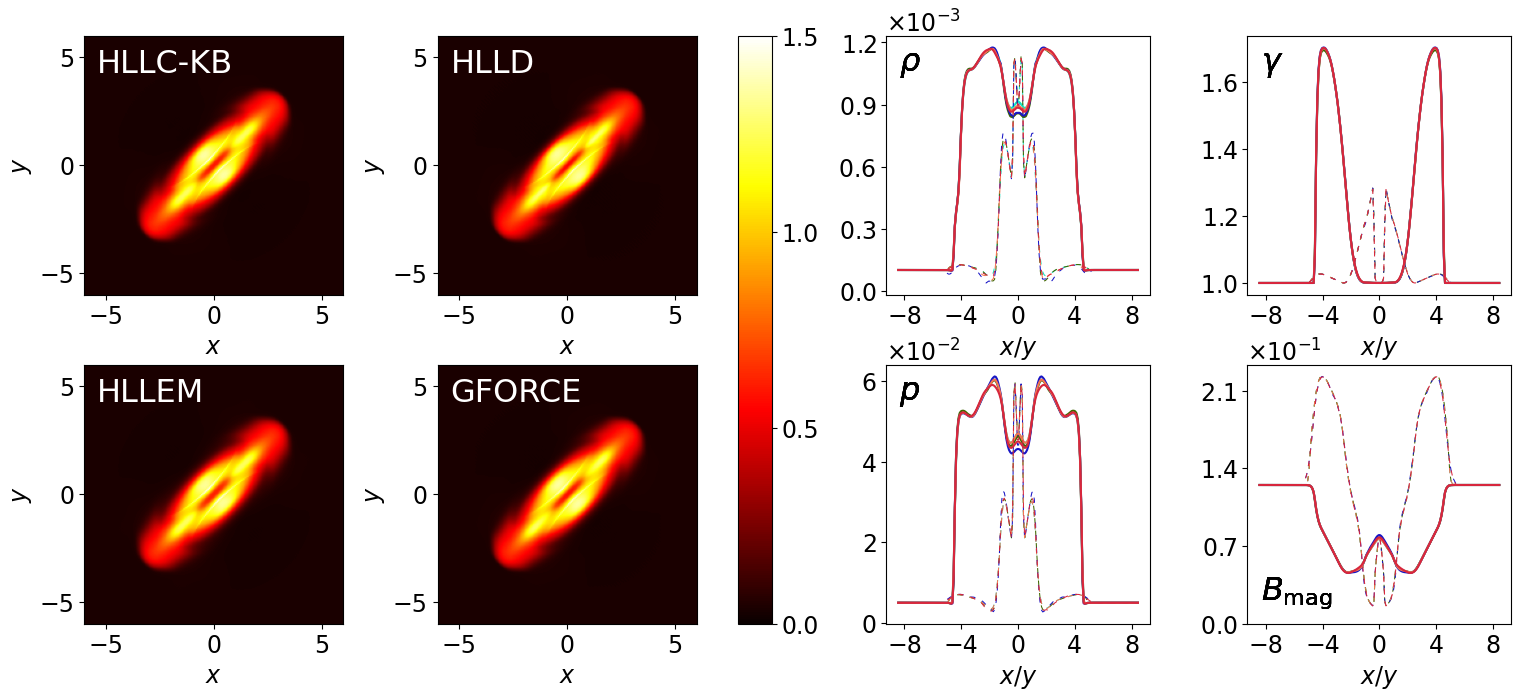}
\footnotesize
\caption{Same as Fig. \ref{fig:BW2D_B0.5_phi0} but for the inclined case ($\phi=45^\circ$).}
\normalsize
\label{fig:BW2D_B0.5_phi45}
\end{figure*}

\begin{figure*}
\centering
\includegraphics[width=0.99\textwidth]{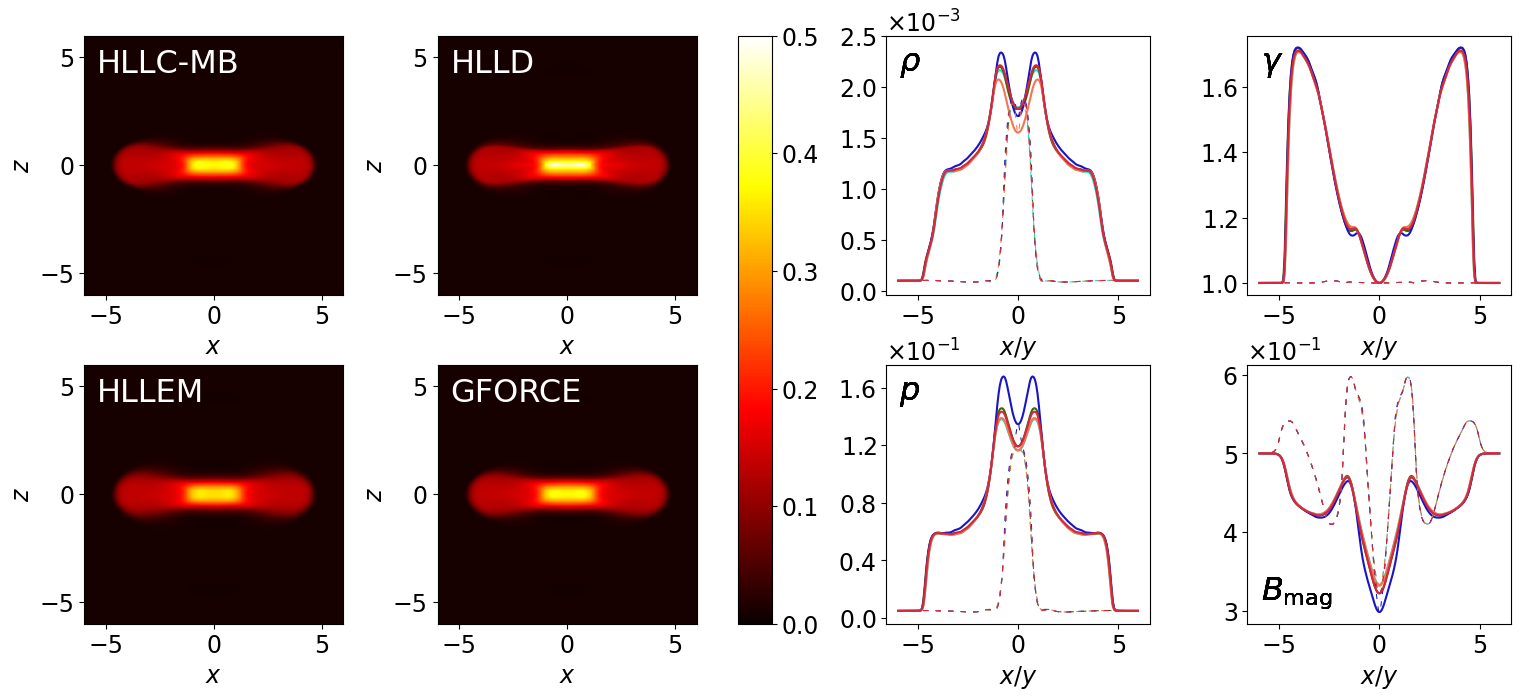}
\footnotesize
\caption{Same as Fig. \ref{fig:BW2D_B0.5_phi0} but for the 3D case and strong magnetization $B_0=1$. 
Colored maps are shown in the $xz$ plane while 1D profiles are taken along the $x$-axis and $y$-axis.
}
\normalsize
\label{fig:BW3D_B1.0_phi0.0}
\end{figure*}

\begin{figure*}
\centering
\includegraphics[width=0.99\textwidth]{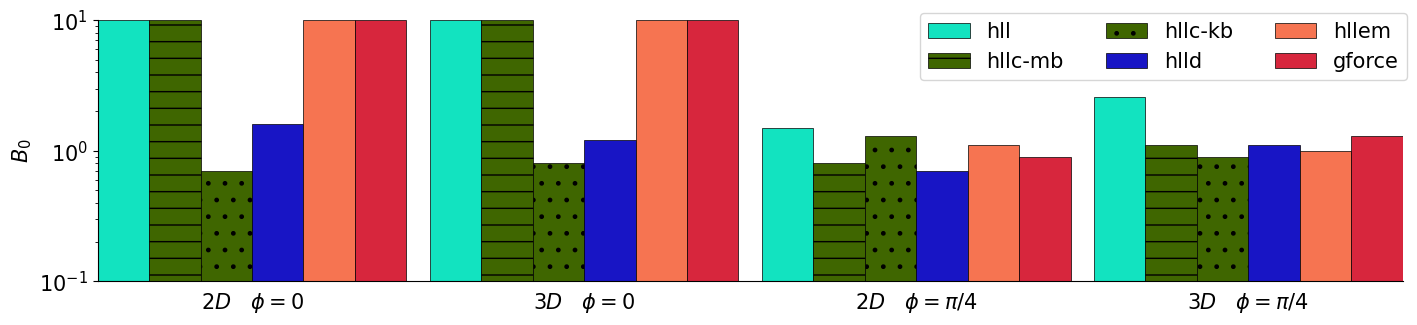}
\footnotesize
\caption{Permitted magnetization values for the blast wave problem.
From left to right the four histograms (2D and 3D with $\phi=0$, 2D and 3D with $\phi=\pi/4$) cover the values of $B_0$ (in the range $[0,10]$) for which numerical integration succeeded.
Each color bar corresponds to a different Riemann solver.
The minmod limiter has been used.
}
\normalsize
\label{fig:BW_breakpoint}
\end{figure*}

Cylindrical and spherical explosions in Cartesian coordinates challenge the robustness of the method and its response to different kinds of degeneracies.

Among the several variants of this problem discussed in the literature \citep[see, for instance,][and reference therein]{dZBL2003, MB2006,dZ_etal2007, BeckStone2011, Marti2015, BalKim2016} here we consider the configuration given by \cite{BeckStone2011}.
In the (original) 2D version of the problem, the computational domain is defined by the square $x,y\in[-6,6]$ initially filled with a uniform ($\rho=10^{-4}$, $ p = 5\cdot10^{-3}$) and static ($\vec{v}=0$) medium.
In 3D, the domain becomes a cube with $z\in[-6,6]$.
A high-pressure region is set up inside the region $r < 0.8$ having $\rho=10^{-2}$, $p=1$, where $r$ is the cylindrical (in 2D) or spherical (in 3D) radius.
The computational domain is threaded with a uniform magnetic field
\begin{equation}
   \vec{B}=B_0\left[\sin\theta\left(
     \cos\phi\hvec{e}_x
    +\sin\phi\hvec{e}_y\right)
    +\cos\theta\hvec{e}_z\right] \,,
\end{equation}
where $\theta$ and $\phi$ are the polar and azimuthal angles, respectively (we set $\theta=\pi/2$).
The grid resolution is fixed to $200^2$ grid zones in 2D and $192^3$ in 3D and computations are carried out until $t=4$ using the ideal EoS with $\Gamma=4/3$.

We begin by showing, in Fig. \ref{fig:BW2D_B0.5_phi0} and \ref{fig:BW2D_B0.5_phi45}, the results of 2D computations using, respectively, $\phi=0$ (grid aligned) and $\phi=\pi/4$ (oblique case) and moderate magnetization $B_0=0.5$.
The left and right halves of the figures include, respectively, a colored map of the plasma $\beta=2p_g/B^2$ (left half) and 1D-profiles along the $x$- and $y$-axis (in the aligned case) or along the two diagonals (in the oblique case).
The explosion is delimited by an outer fast forward shock and the presence of a magnetic field makes the propagation anisotropic by compressing the gas in the direction parallel to the field. 
In the perpendicular direction the outer fast shock becomes magnetically dominated with very weak compression. 
Results between different solvers are very similar and the salient features of the solution are confirmed  also in the oblique case.

In 3D and for stronger magnetization ($B_0=1$),  differences are slightly more emphasized around the center where less diffusive solvers such as HLLD and GFORCE yield larger density and pressure peaks and smaller magnetic energies, see Fig. \ref{fig:BW3D_B1.0_phi0.0}.
We point out that the HLLD solver and the two flavors of HLLC could not successfully complete the 3D case with $B_0=1$ without enabling the corresponding \quotes{failsafe} switches to HLL (see the discussion in sections \S \ref{sec:HLLC_MB}, \S\ref{sec:HLLC_KB} and \S\ref{sec:HLLD}).

It should be clear by now that the stability of the computations crucially depends on the chosen solver. 
Fig. \ref{fig:BW_breakpoint} reports the allowed range of magnetization values  (above which computation breaks down) for different Riemann solvers using different inclinations in the $x-y$ plane ($\phi = 0$ and $\phi=45^\circ$) in 2D as well as in 3D.
The histograms have been obtained by increasing $B_0$ in steps of $0.1$ in the range $[0,10]$ for each computations. 
Overall, larger magnetizations are attained for grid-aligned configurations ($\phi=0$) in both 2D and 3D while the oblique cases appear to be more stringent in terms of stability. 
In the former case, HLL, HLLEM and GFORCE (with $\omega=1/2$) yield the most robust results.
In the oblique cases, however, the maximum permitted values decrease to values of order unity.
The HLLC-MB solver seems to be more robust than the KB version for grid-aligned configurations while it becomes comparable for $\phi=\pi/4$.
We point out that the limits have been obtained by quitting the computation at the first failure of the conservative to primitive inversion scheme.
Larger values may be possible by applying corrections to energy and/or momentum \citep[see, e.g., the work of][]{MB2006, BeckStone2011, Marti2015}.

The CPU time required by the different Riemann solvers on this particular test were found to be $t_{\rm hll}:t_{\rm hllc-MB}:t_{\rm gforce}:t_{\rm hllc-KB}:t_{\rm hlld}:t_{\rm hllem} = 1:1.07:1.43:1.47:1.72:2.43$ (the CPU time for the HLLC solvers is computed in the case with $B_0=0.1$).

\subsection{Kelvin-Helmholtz Instability}
%

\begin{figure}
\centering
\includegraphics[width=0.49\textwidth]{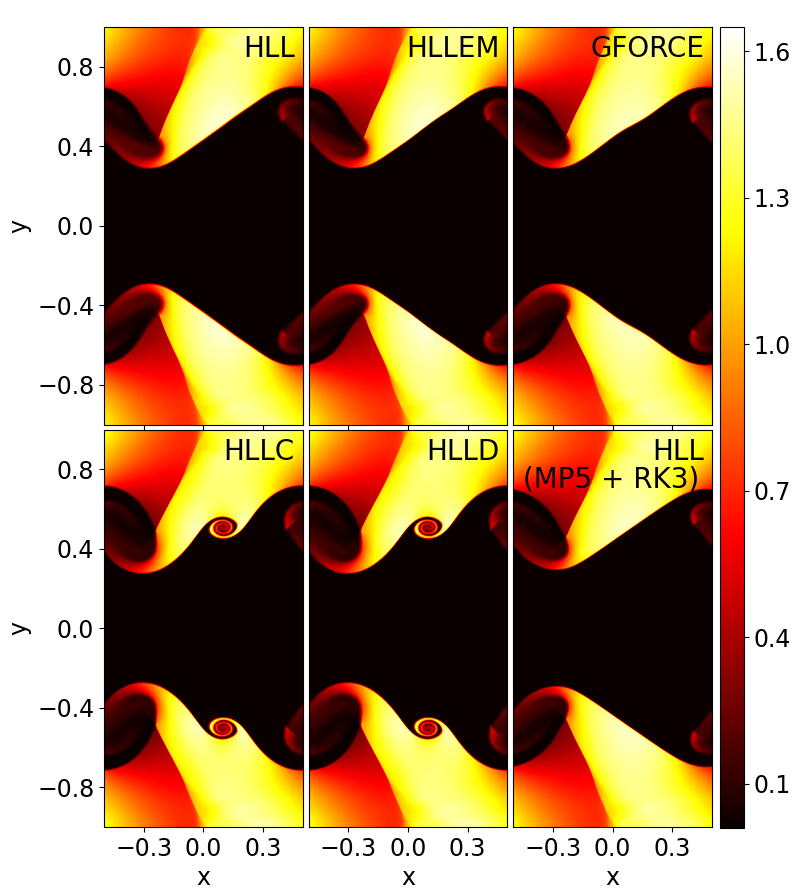}
\footnotesize
\caption{Density distribution of the Kelvin-Helmoltz instability test problem at $t = 3$ with different Riemann solvers. All the runs have been performed with $512\times1024$ grid cells.
}
\normalsize
\label{fig:KHId}
\end{figure}

As a final test we choose the 2D Kelvin-Helmholtz instability (KHI) using the configuration of \citet{BeckStone2011}.
The initial shear velocity is given by
\begin{equation}
    v^x = \sign(y)v_{\rm sh}\tanh\left[\DS\frac{2y-\sign(y)}{2a}\right] \,,
\end{equation}
where $a = 0.01$ represents the thickness of the shear layer and $v_{\rm sh} = 0.5$.
The shear layer is perturbed by a non-zero $y$-component of the velocity:

\begin{equation}
    v^y = \sign(y)A_0v_{\rm sh}\sin(2\pi x)\exp\left[-\left(\DS\frac{2y-\sign(y)}{2\sigma}\right)^2\right] \,,
\end{equation}
where $A_0=0.1$ is the amplitude of the perturbation and $\sigma = 0.1$ is the perturbation length-scale.
We set an uniform initial pressure $p = 1.0$ and employ the ideal EoS with adiabatic exponent $\Gamma = 4/3$, while the magnetic field is non-zero only in the x-direction ${\vec B} = (10^{-3},0,0)$.
Finally, the density distribution is set as:
\begin{equation}
    \rho = 
    \DS\frac{1}{2}(\rho_l + \rho_h) + \DS\frac{1}{2}(\rho_h - \rho_l)\DS\frac{v^x}{v_{\rm sh}} \,,
\end{equation}
with $\rho_h = 1.0$ and $\rho_l = 0.01$.
The Cartesian domain has extension of $x\in[-0.5,0.5]$, $y\in[-1.0,1.0]$ with periodic boundary conditions applied in all directions.
We use a nominal resolution of $512\times1024$ grid zones and evolve the system until $t=3$.
Lower resolutions ($128\times256$ and $256\times512$) have been employed for convergence purposes.

Our results confirm and extend those obtained by \citet{BeckStone2011}, namely, that the choice of the Riemann solver plays a crucial role in its ability to capture the turbulence at smaller scales leads to an increase in the effective resolution.
The density maps shown in Fig. \ref{fig:KHId} show, in fact, that only the HLLC and HLLD solvers are able to capture small scale structure (i.e. the secondary vortexes at $t = 3$) while, on the contrary, the remaining solvers (HLL, GFORCE and HLLEM) disclose a lesser amount of substructure and a larger amount of numerical diffusion, even at very high resolution (not shown here).

The same setup has been tested also employing the HLL Riemann solver and a higher order scheme (in particular, a parabolic reconstruction and a $3^{\rm rd}$-order time integration Runge-Kutta scheme have been adopted). As shown in the bottom right panel of Fig. \ref{fig:KHId}, the secondary vortexes are not developing.

Interestingly, the differences between the HLL and the HLLEM solver are almost negligible regardless of the resolution, even though the HLLEM solver is designed to preserve the contact wave.

In order to explain this apparently unexpected behavior, we first observe that this problem is i) only weakly magnetized ($\beta\sim10^5$) and ii) strictly two-dimensional (no z component is present).
These conditions imply that slow waves become almost degenerate on the contact mode, while Alfv{\'e}n waves are not present in the solution.
Thus only 3 (out of 5) waves can be accounted for by the HLLEM solver: two outermost acoustic waves and the middle contact mode describing a density jump.
When $B_x\to0$, however, the middle wave is best identified as a tangential discontinuity, carrying jumps in the transverse vector components as well.
These variations are crucial in the vortex formation process but they cannot be described and are thus smoothed out by the HLLEM solver.
On the contrary, both HLLC and HLLD solvers are able to capture the discontinuities in the transverse components of the velocity, even if none of them is specifically designed to fully capture slow waves.
Both solvers, in fact, are able to \quotes{detect} a transverse velocity jump\footnote{For HLLC-MB, this statement holds in the $B_x\to0$ limit.} since this is inherently part of the nonlinear solution process.

\begin{figure}
\centering
\includegraphics[width=0.49\textwidth]{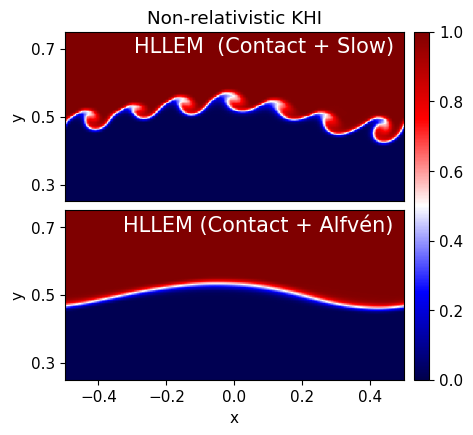}
\footnotesize
\caption{Density maps for the KHI instability at $t = 1.5$ in the non-relativistic case using the 5-waves HLLEM solver including contact + slow waves (top panel) and including contact + Alfv{\'e}n waves.
}
\normalsize
\label{fig:KHIMHD}
\end{figure}
In order to prove our statement, we now show that restoration of the slow modes in the HLLEM Riemann solver is decisive in resolving small-scale structure.
We demonstrate this by performing the same computation in the non-relativistic regime (MHD), since this sensibly reduces the required computational time (as shown in \citealt{Anton_etal2010}). 
Two sets of solvers have been considered: in the first case (top panel in Fig. \ref{fig:KHIMHD}) the HLLEM solver is designed to capture contact and slow modes, while in the second case (bottom panel of the same figure), the HLLEM solver resolves contact and Alfv\'en waves.
A comparison between the two panels in Fig. \ref{fig:KHIMHD} clearly reveals that the former is able to resolve multiple secondary vortices across the shear layer while the latter completely smooths them out.

\begin{figure}
\centering
\includegraphics[width=0.49\textwidth]{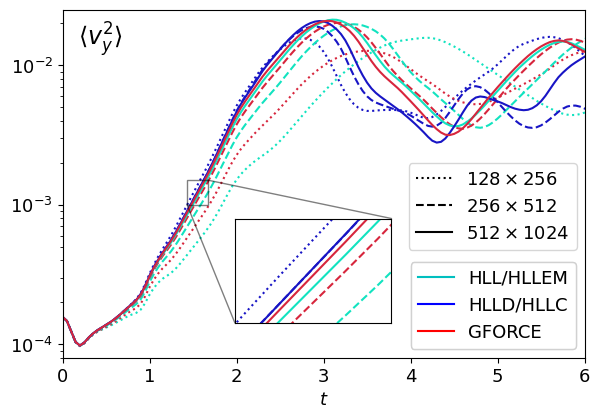}
\footnotesize
\caption{Growth rate, defined as $\langle|v_y|^2\rangle$ computed using different solvers at different resolutions.
Since HLL and HLLEM show the same growth rate, they have been represented by a single line.
We applied the same strategy also for the HLLC/HLLD solvers.
}
\normalsize
\label{fig:KHIg}
\end{figure}

Finally, we provide in Fig. \ref{fig:KHIg} a measure of the instability growth rate through the volume-integrated transverse velocity squared, at different resolutions $\langle|v_y|^2\rangle$. 
While the HLLC and the HLLD solvers converge almost immediately, the GFORCE, HLL and HLLEM solvers achieve complete convergence only at larger resolutions.
In spite of this, the GFORCE scheme approach the nominal growth rate at a somewhat faster rate when compare to HLL or HLLEM.

%% file: summary.tex
\section{Summary}
\label{sec::Summary}
%
%
%

A comparison between several non-linear approximate Riemann, namely, HLL, HLLC, HLLD, HLLEM and GFORCE, has been presented through a series of 1, 2 and 3D numerical tests, in order to assess their efficiency, stability and robustness.
Our conclusions, based on the employment of second-order reconstruction schemes, can be summarized as follows:

\begin{itemize}
\item Owing to its ability to approximate the Riemann fan structure by including rotational and contact discontinuities, the HLLD solver of \citet{MUB2009} is able to achieve the best results in terms of accuracy.
Despite being more computationally expensive than more diffusive Riemann solvers, its ability to converge at lower resolution allows comparable accuracies to be achieved with a reduced number of grid cells (e.g., $256\times512$ vs $512\times1024$ required by HLL, see the the Kelvin-Helmoltz instability problem).
On the other hand, because of its complex and iterative character, this solver may not be a robust option for strong magnetizations.

\item While the HLL Riemann solver showed great performances in terms of stability and computational efficiency, its inability of resolving any internal wave of the Riemann fan lead to a very diffusive behavior in all of the presented tests.

\item The HLLC Riemann solver showed dissipation properties intermediate between the HLLD and the HLL formulation.
Since several approaches have been developed through the years, the approaches of \citet{MB2006} and of \citet{BalKim2016} have been compared.
The former formulation (HLLC-MB) showed a better performance in terms of computational efficiency, since it does not require any iterative cycle.
On the contrary, the second approach (HLLC-KB) involves the solution of couples systems of nonlinear equations and it is thus more computational intensive.
In addition we found that the HLLC-KB solver is not fully consistent with the integral form of the conservation law, failing to satisfy some jump condition across the contact mode.

\item The GFORCE Riemann solver shows accuracy comparable (or slightly inferior) to the HLLC approach.
Its increased stability properties, which are intermediate between the HLL and the HLLD approach, makes it a valid robust alternative when the HLLD Riemann solver becomes brittle.
The solver is non-iterative and it requires one additional conversion from conservative to primitive variables slowing down the flux computation by approximately $50\%$ when compared to the HLL solver. 
In the presence of strong shock and/or magnetizations, the GFORCE should be reduced, in our experience, to the FORCE flux by tuning the parameter $\omega_g=1/2$ for safety purposes.

\item the HLLEM formulation, despite its ability of resolving the contact and rotational discontinuities, has often shown poor accuracy and numerical dissipation comparable to the HLL formulation.
Better performances can be obtained when Alfv\'en waves are predominant, although not superior than the HLLD formulation.
Since its intermediate eigenstructure is built on top of the HLL solver, its stability properties are better than other less diffusive Riemann solvers.
On the other hand, its large computational cost (related to the computation of left and right eigenvectors) does not make it - in our opinion - an efficient and valid alternative in the context of relativistic MHD, although its performance may improve for those systems where conserved eigenvectors are easier to compute (e.g. non relativistic MHD equations).
\end{itemize}

%% file: appendix.tex
\section{A note on the eigenvectors of RMHD}
\label{sec:Appeigenvec}
%

The eigenstructure of the RMHD equations has been studied by \citet{Anile_Pennisi1987,Anile2005},
and rewritten by \citet{Komissarov1999,Balsara2001,Anton_etal2010} in a more suited way for the numerical schemes.
Our method of solution follows the approach of \citet{Anton_etal2010}, although the computation of the left
eigenvectors slightly differs from their approach.
For the sake of clarity, we summarize here the pertinent formulas.
The most convenient way to compute the left and right eigenvectors is to use the so-called covariant variables
$\tilde{\cU} = (u^\mu,b^\mu,p,s)^T$.
The eigenvector problem becomes 
\begin{equation}
\label{eq::leftrighteigen}
({\cal A}^\mu\phi_\mu)\vec{\tilde{r}} = 0 \qquad \vec{\tilde{l}_0}({\cal A}^\mu\phi_\mu) = 0 
\end{equation}
The vector $\phi_\mu = (-\lambda,1,0,0)$ describes the normal to the characteristic hypersurface, while the matrices ${\cal A}^\mu$ are defined by
\begin{equation}
\label{eq::eigenvecmat}
{\cal A}^\mu = \left(
 \begin{array}{cccc}
   {w_T}u^\mu\delta^\alpha_\beta & -b^\mu\delta^\alpha_\beta + P^{\alpha\mu}b_\beta & l^{\alpha\mu} & 0^\alpha \\ \noalign{\medskip}
  b^\mu\delta^\alpha_\beta & -u^\mu\delta^\alpha_\beta & f^{\mu\alpha} & 0^\alpha \\ \noalign{\medskip}
  \rho h\delta^\mu_\beta & 0_\beta & u^\mu/c_s^2 & 0 \\ \noalign{\medskip}
  0_\beta & 0_\beta & 0 & u^\mu
 \end{array}
\right)
\end{equation}
where the index $\alpha = [0,1,2,3]$ indicates the rows and the index $\beta = [0,1,2,3]$ indicates the columns.
The quantities introduced in Eq. \ref{eq::eigenvecmat} are 
$P^{\alpha\mu}_\beta = {\tens \eta}^{\alpha\mu} + 2u^\alpha u^\mu$, 
$l^{\alpha\mu} = (\rho h{\tens \eta}^{\alpha\mu} + (\rho h - b^2/c_s^2)u^\alpha u^\mu)/(\rho h)$, 
$f^{\alpha\mu} = (u^\mu b^\alpha/c_s^2 - u^\alpha b^\mu)/(\rho h)$, while $c_s$ is the sound speed.

As pointed by \citet{Koldoba_etal2002,Anton_etal2010}, the orthonormalization of the eigenvectors is provided by
\begin{equation}
 \vec{\tilde{l}_0}(\lambda_1){\cal A}^0\vec{\tilde{r}}(\lambda_2) = \vec{\tilde{l}}(\lambda_1)\vec{\tilde{r}}(\lambda_2) = \delta^{\lambda_1}_{\lambda_2}
\end{equation}
Because of the degeneracies of the RMHD, we renormalized the left and right eigenvectors as already done by \citet{Anton_etal2010}.
We start with the right eigenvector associated to the entropy wave,
\begin{equation}
\vec{\tilde{r}}_e = (0^\alpha,0^\alpha,0,1)^T.
\end{equation}
In order to compute the right Alfv\'en eigenvectors we need some intermediate quantities, as
\begin{equation}
\begin{array}{lcr}
\alpha_1^\mu & = & \gamma(v^z,\lambda_av^z,0,1-\lambda_av^x) \\ \noalign{\medskip}
\alpha_2^\mu & = & -\gamma(v^y,\lambda_av^y,1-\lambda_av^x,0), 
\end{array}
\end{equation}
and
\begin{equation}
\begin{array}{lcl}
 g_1 = \DS\frac{1}{\gamma}\left(B^y + \DS\frac{\lambda_av^y}{1-\lambda_av^x}B^x\right) \\ \noalign{\medskip}
 g_2 = \DS\frac{1}{\gamma}\left(B^z + \DS\frac{\lambda_av^z}{1-\lambda_av^x}B^x\right)
\end{array}
\end{equation}
where, if $g_1 = g_2 = 0$, we follow the prescription $g_1 = g_2 = 1$.
The explicit form of the right Alfv\'en eigenvectors becomes
\begin{equation}
 \vec{\tilde{r}}_{a,\pm} = (f_1\alpha_1^\mu + f_2\alpha_2^\mu, \mp\sqrt {w_T}(f_1\alpha_1^\mu + f_2\alpha_2^\mu),0,0)^T,
\end{equation}
where
\begin{equation}
 f_{1,2} = \frac{g_{1,2}}{\sqrt{g_1^2+g_2^2}}.
\end{equation}
The normalized left eigenvectors in covariant variable are computed using Eq. \ref{eq::leftrighteigen},
which leads to
\begin{equation}
 \vec{\tilde{l}}_e = (0^\alpha,0^\alpha,0,1)
\end{equation}
for the entropy eigenvector, and
\begin{equation}
\vec{\tilde{l}}_{a,\pm} = N\left(
\begin{array}{c}
( {w_T}\gamma \pm b^0\sqrt {w_T})(f_1\alpha_{1\mu} + f_2\alpha_{2\mu}) \\
-(b^0 \pm\sqrt {w_T}\gamma)(f_1\alpha_{1\mu} + f_2\alpha_{2\mu}) + (f_1\alpha_1^0+f_2\alpha_2^0)b_\mu \\
f_1\alpha_1^0+f_2\alpha_2^0 \\
0
\end{array}
\right)^T
\end{equation}
for the Alfv\'en eigenvectors.
The renormalization factor $N$ takes the form
\begin{equation}
N = \DS\frac{\sqrt {w_T}}{g_1^2+g_2^2}(N_1 + N_2 + N_3),
\end{equation}
where
\begin{equation}
\left\{
\begin{array}{lcl}
 N_1 & = & (B^zv^y - B^yv^z)^2[2(\lambda^2-1)\sqrt {w_T}\gamma+b^0\sqrt {w_T}(2\lambda^2-1)\mp\lambda b^x] \\ \noalign{\medskip}
 N_2 & = & 2(\sqrt {w_T}\gamma\pm b^0)(\gamma-\lambda u^x)^2(g_1^2+g_2^2) \\ \noalign{\medskip}
 N_3 & = & (B^yv^z - B^zv^y)(\gamma-\lambda u^x)(b^zg_1-b^yg_2) 
\end{array}
\right..
\end{equation}
This normalization is well defined through the RMHD degeneracies.

In order to include the entropy and Alfv\'en waves in the HLLEM solver we have to compute
the normalized eigenvectors in conserved variables.
The conserved eigenvectors are computed as follows:
\begin{equation}
{\vec R} = \left(\DS\pd{\cU}{\vec{\tilde{\cU}}}\right)\vec{\tilde{r}} \qquad {\vec L} = \vec{\tilde{l}}\left(\DS\pd{\vec{\tilde{\cU}}}{\cU}\right)
\end{equation}
The transformation matrix for the right eigenvectors has a straightforward explicit form:
\begin{equation}
    \left(\pd{\cU}{\vec{\tilde{\cU}}}\right) = 
    \left(\begin{array}{cccccc}
    \rho & 0^j  & 0 & 0^j & \rho_p\gamma & \rho_s\gamma \\
     {w_T}u^i &  {w_T}\gamma\delta^{ij} & A^i & M^{ij} & w_p\gamma u^i & w_s\gamma u^i \\
    b^i & -b^0\delta^{ij} & -u^i & \gamma\delta^{ij} & 0 & 0\\
    2{w_T}\gamma - \rho & 0^j & F & C^j & G & w_s
    \end{array}\right) 
\end{equation}
where the intermediate quantities are

\begin{equation}
 \arraycolsep=1.4pt
 \begin{array}{llllll}
   M^{ij}    &=&  2b^j\gamma u^i - b^0\delta^{ij} 
 & \qquad A^i &=& -2b^0\gamma u^i - b^i 
 \\ \noalign{\smallskip}
    C^j    &=&  2b^j\gamma^2 - b^j 
 & \qquad F &=& -2b^0\gamma^2 - b^0 
 \\ \noalign{\smallskip}
   G       &=& w_p\gamma^2 - 1 - \rho_p\gamma
 \end{array}
\end{equation}
while the partial derivatives are written in a more compact form
\begin{equation}
\begin{array}{l}
 \rho_s = \left(\DS\pd{\rho}{s}\right)_p = -\DS\frac{\rho}{s\Gamma}  \quad w_s = \left(\DS\pd{\rho h}{s}\right)_p = -\DS\frac{\rho}{s\Gamma} \\ \noalign{\medskip}
 \rho_p = \left(\DS\pd{\rho}{p}\right)_s = \DS\frac{\rho}{\Gamma p}  \quad w_p = \left(\DS\pd{\rho h}{p}\right)_s = \DS\frac{\rho}{\Gamma p} + \DS\frac{\Gamma}{\Gamma - 1},
\end{array}
\end{equation}
assuming an ideal equation of state.
The conversion to the conserved variables yields
\begin{equation}
{\vec R}_e = -\DS\frac{D}{s\Gamma}(1,u^x,u^y,u^z,\gamma - 1,0,0,0)^T
\end{equation}
for the entropic eigenvector, and
\begin{equation}
{\vec R}_{a,\pm} = f_1{\vec V}_{a,1,\pm} - f_2{\vec V}_{a,2,\pm}
\end{equation}
for the Alfv\'en eigenvectors, where
\begin{equation}
{\vec V}_{a,1,\pm} = 
\left(
    \begin{array}{c}
    \rho u^z \\
    2u^z({w_T}u^x \pm\sqrt{w_T}b^x) \\
    {w_T}u^yu^z \pm\sqrt{w_T}b^yu^z \\
    {w_T}[\gamma^2 + (u^z)^2 - (u^x)^2]\pm\sqrt{w_T}(b^zu^z + b^0\gamma - b^xu^x) \\
    0 \\
    b^yu^z \pm \sqrt{w_T}u^yu^z \\
    -b^yu^y \mp \sqrt{w_T}[1 + (u^y)^2] \\
     2u^z({w_T}\gamma\pm\sqrt{w_T}b^0) - \rho u^z
    \end{array}\right)
\end{equation}
and
\begin{equation}
{\vec V}_{a,2,\pm} = 
  \left(
    \begin{array}{c}
    \rho u^y \\
    2u^y({w_T}u^x\pm\sqrt{w_T}b^x) \\
    {w_T}[\gamma^2 + (u^y)^2 - (u^x)^2]\pm\sqrt{w_T}(b^yu^y + b^0\gamma - b^xu^x) \\
    {w_T}u^yu^z \pm{w_T}b^zu^y \\
    0 \\
    -b^zu^z \pm\sqrt{w_T}[1+(u^z)^2] \\
    b^zu^y\pm{w_T}u^yu^z \\
    2u^y({w_T}\gamma \pm\sqrt{w_T}b^0) - \rho u^y
    \end{array}\right)
\end{equation}
The computation of the transformation matrix is made, as in \citet{Anton_etal2010}, in two steps.
The first step is to convert the eigenvectors in primitive variables
$\vec{\bar{\cV}} = (u^x,u^y,u^z,b^x,b^y,b^z,p,\rho)$,
\begin{equation}
 \vec{l} = \vec{\tilde{l}}\left(\DS\pd{\vec{\tilde{\cU}}}{\vec{\bar{\cV}}}\right),
\end{equation}
while, in the second step we recover directly the scalar product ${\vec L}_*\cdot(\cU_R - \cU_L)$ , which is computed taking the scalar product between the primitive eigenvectors $\vec{\tilde{l}}_*$ and the solution of the linear system
\begin{equation}
 \left(\DS\pd{\cU}{\bar{\cV}}\right){\vec X} = \cU_R - \cU_L,
\end{equation}
where ${\vec X}$ is the unknown vector.
The first transformation matrix has the form
\begin{equation}
 \left(\DS\pd{\tilde{\cU}}{\bar{\cV}}\right) = 
 \left(\begin{array}{cccc}
   v^j & 0^j & 0 & 0 \\ \noalign{\medskip}
   \delta^{ij} & 0^{ij} & 0 & 0 \\ \noalign{\medskip}
   B^j & u^j & 0 & 0 \\ \noalign{\medskip} 
   \DS\pd{b^i}{u^j} & \DS\pd{b^i}{B^j} & 0 & 0 \\ \noalign{\medskip}
   0^j & 0^j & 1 & 0 \\ \noalign{\medskip}
   0^j & 0^j & \left(\DS\pd{s}{p}\right)_\rho & \left(\DS\pd{s}{\rho}\right)_p
 \end{array}\right) 
\end{equation}
where
\begin{equation}
\begin{array}{lcl}
\DS\pd{b^j}{u^i} & = & v^iB^j - B^iv^j\gamma^{-2} - ({\vec v}\cdot{\vec B})(v^iv^j - \delta^{ij}) \\ \noalign{\medskip}
\DS\pd{b^j}{B^i} & = & \gamma^{-1}(u^iu^j + \delta^{ij})
\end{array},
\end{equation}
and
\begin{equation}
\left(\DS\pd{s}{\rho}\right)_p = \DS\frac{s}{p} \quad \left(\DS\pd{s}{p}\right)_\rho = -\DS\frac{s\Gamma}{\rho}.
\end{equation}
A difference between our approach and the one of \citet{Anton_etal2010} is that, since the conversion matrix
is less straightforward, we do not provide an analytical expression for the left eigenvectors in primitive variables.
On the other hand, this approach, since it converts immediately from the covariant magnetic field to the laboratory magnetic field,
the latter step is much easier to compute.

The explicit form of the latter transformation matrix is
\begin{equation}
 \left(\DS\pd{\cU}{\bar{\cV}}\right) = 
 \left(\begin{array}{cccc}
  \rho v^j & 0^j & 0 & \gamma \\  \noalign{\medskip}
  \DS\pd{S^i}{u^j} & \DS\pd{S^i}{B^j} & \DS\frac{\Gamma}{\Gamma - 1}\gamma u^i & \gamma u^i \\  \noalign{\medskip}
  0^{ij} & \delta^{ij} & 0^i & 0^i \\  \noalign{\medskip}
  \DS\pd{E}{u^j} & \DS\pd{E}{B^j} & \DS\frac{\Gamma}{\Gamma - 1}\gamma^2 - 1 & \gamma(\gamma - 1)
 \end{array}\right),
\end{equation}
where the partial derivatives are
\begin{equation}
\begin{array}{lcl}
\DS\pd{S^i}{u^j} & = & (\rho h - \DS\frac{B^2}{\gamma^2})v^iu^j - \DS\frac{B^iB^j}{\gamma} + \DS\frac{B^iv^j}{\gamma}({\vec v}\cdot{\vec B}) + (Dh + \DS\frac{B^2}{\gamma})\delta^{ij} \\ \noalign{\medskip}
\DS\pd{S^i}{B^j} & = & 2v^iB^j -B^iv^j - ({\vec v}\cdot{\vec B})\delta^{ij} \\ \noalign{\medskip}
\DS\pd{E}{u^j}   & = & 2u^j\rho h - \rho v^j + \DS\frac{v^jB^2 - B^j({\vec v}\cdot{\vec B})}{\gamma} - [v^2B^2 - ({\vec v}\cdot{\vec B})^2]\DS\frac{v^j}{\gamma} \\ \noalign{\medskip}
\DS\pd{E}{B^j}   & = & B^j(1 + v^2) - v^j({\vec v}\cdot{\vec B})
\end{array}.
\end{equation}
We point out that the system has a trivial solution in the magnetic field components, therefore it can be reduced to 5 unknown values in order to increase its speed and performance.
Although the last two steps are performed numerically, the orthonormalization of the conserved eigenvectors is preserved up to machine accuracy.